\documentclass[epj]{svjour}
\usepackage{amsfonts,amssymb,amsmath}
\usepackage{dcolumn,epsfig,color}

\newcommand{\fet}[1]{\mbox{\boldmath $#1$}}

\begin{document}

\title{Dependence of the triple-alpha process on the fundamental
constants of nature}

\author{Evgeny Epelbaum\inst{1} 
\and Hermann~Krebs\inst{1} 
\and Timo A. L\"{a}hde \inst{2} 
\and Dean~Lee\inst{3} 
\and Ulf-G.~Mei{\ss}ner\inst{2,4,5} 
%
}                     
%
%
\institute{Institut~f\"{u}r~Theoretische~Physik~II,~Ruhr-Universit\"{a}t~Bochum,~D-44870~Bochum,~Germany
\and Institut~f\"{u}r~Kernphysik,~Institute~for~Advanced~Simulation,
J\"{u}lich~Center~for~Hadron~Physics, \\ 
Forschungszentrum~J\"{u}lich, D-52425~J\"{u}lich,~Germany
\and Department~of~Physics,~North~Carolina~State~University,~Raleigh,~NC~27695,~USA
\and JARA -- High~Performance~Computing,~Forschungszentrum~J\"{u}lich,~D-52425~J\"{u}lich,~Germany
\and Helmholtz-Institut~f\"{u}r~Strahlen-~und~Kernphysik~and~Bethe~Center~for
Theoretical~Physics, Universit\"{a}t~Bonn, \\ D-53115~Bonn,~Germany
}
\date{Received: date / Revised version: date}
%

\abstract{
We present an {\it ab initio} calculation of the quark mass dependence of the ground 
state energies of $^4$He, $^8$Be and $^{12}$C, and of the energy of the Hoyle
state in $^{12}$C. These investigations are performed within the framework of
lattice chiral Effective Field Theory.
We address the sensitivity of the production rate of carbon and oxygen in red giant stars to the
fundamental constants of nature by considering the impact of variations in the light quark
masses and the electromagnetic fine-structure constant on the reaction rate of the triple-alpha process.
As carbon and oxygen are essential to life as we know it, we
also discuss the implications of our findings for an anthropic view of the Universe.
We find strong evidence that the physics of the triple-alpha process is driven by alpha 
clustering, and that shifts in the fundamental parameters at the $\simeq 2 - 3$\% level are unlikely to be 
detrimental to the development of life. Tolerance against much larger changes cannot be ruled out at 
present, given the relatively limited knowledge of the quark mass dependence of the two-nucleon $S$--wave 
scattering parameters. Lattice QCD is expected to provide refined
estimates of the scattering parameters in the  future.
} 

\maketitle


\section{Introduction \label{intro}}

The ``Hoyle state'' is an excited state of $^{12}$C with quantum numbers $J^p_{} = 0^+$ 
just above the $^{8}$Be--$\alpha$ threshold. The existence of such a state was  predicted by 
Hoyle in order to explain the observed abundance of $^{12}$C, which is produced
during helium burning in red giant stars via the so-called triple-alpha process. In this two-step process, 
two $^4$He nuclei first combine to form the unstable, but relatively long-lived (under the conditions
prevalent in the cores of red giant stars) $^8$Be nucleus. This resonance must then combine with a third
alpha particle in order to generate carbon.

However, the fact that this process cannot by itself explain the observed
abundance of $^{12}$C in the Universe motivated Hoyle in 1954 to propose the
existence 
of an excited $0^+$ state of $^{12}$C, 
just above the $^8$Be--$\alpha$ threshold. Such a resonant enhancement could 
then provide a sufficiently high
rate of production of $^{12}$C to account for the observed
abundance~\cite{1}. Soon afterwards, the predicted
state was detected at Caltech~\cite{2,3}, and the modern value for its energy
is $\varepsilon = 379.47(18)$~keV  relative to the $3\alpha$ threshold, while
the total and radiative widths are $\Gamma_{\rm tot}^{} = 8.3(1.0)$~eV and
$\Gamma_{\gamma}^{} = 3.7(5)$~meV.

While the Hoyle state dramatically increases the reaction rate of the
triple-alpha process, the resulting enhancement
is also extremely sensitive to the exact value of $\varepsilon$, which is
therefore the principal control parameter of
this reaction. As the Hoyle state is crucial to the formation of elements
essential to life as we know it, this state
has been nicknamed the ``level of life''~\cite{Linde} 
Thus, the
Hoyle state is often viewed as a prime manifestation of the anthropic
principle, which states that the observable 
values of the fundamental physical and cosmological parameters are restricted
by the requirement that life can form to 
determine them, and that the Universe be old enough for that to
occur~\cite{6,7}. See, however, Ref.~\cite{Kragh} for a thorough historical
discussion of the Hoyle state in view of the anthropic principle.
We remark that in the context of cosmology and string 
theory, the anthropic principle and its consequences have had a significant influence
(see {\it e.g.} Refs.~\cite{Weinberg:1987dv,Susskind:2003kw}). 

The impact of changes in the energy of the Hoyle state on the synthesis of
$^{12}$C and $^{16}$O in red giant stars
has been investigated in several numerical studies that make use of highly 
sophisticated stellar evolution models.
Livio {\it et al.}~\cite{Livio} modified the value of $\varepsilon$ by hand
and studied the triple-alpha process in the core 
and shell helium burning up to the asymptotic giant branch stage in the 
stellar evolution. These calculations have been
refined by Oberhummer {\it et al.}, who concluded that the production of either $^{12}$C or $^{16}$O
becomes strongly suppressed for changes larger than $\delta(\varepsilon) 
\simeq \pm 100$~keV in the position of the Hoyle state~\cite{Oberhummer_astro}.
In essence, if $\varepsilon$ is lowered too much, the triple-alpha process 
ignites at a significantly lower stellar core
temperature, and hence not much energy is available for the process $^{12}$C +
$^{4}$He $\to$ $^{16}$O + $\gamma$. Conversely, if $\varepsilon$ is raised too
much, the triple-alpha process ignites at a much higher core temperature, and hence most of
the $^{12}$C formed is immediately converted into $^{16}$O and $^{20}$Ne already before the conclusion of core
He burning. However, since a $\simeq 100$~keV change in $\varepsilon$ could
still be tolerated, which is a 25\% modification, the degree of
fine-tuning was revealed to not be as severe as was first believed \cite{WeinbergFacing}.

In addition to these {\it ad hoc} changes in $\varepsilon$, a more microscopic 
calculation was performed by Oberhummer {\it et
  al.}~\cite{Oberhummer:2000mn,Oberhummer:2000zj} 
in terms of a nuclear cluster model based on a simple 
nucleon-nucleon (NN) interaction with inclusion of electromagnetic (EM)
effects. This NN interaction was formulated in terms of one 
strength parameter adjusted to give a fair description of $\alpha$--$\alpha$
scattering and the spectrum of $^{12}$C. By modifying 
this coupling strength and the EM fine structure constant $\alpha_{\rm  em}^{}$, 
the effect on the stellar production of $^{12}$C and 
$^{16}$O was analyzed. Within such a model, an adequate 
amount of $^{12}$C and $^{16}$O was produced within a relatively narrow window of 
$\simeq 0.5$\% around the observed strong force and of $\simeq 4$\% around the 
observed strength of the EM interaction. For larger
changes, the stellar production of carbon and/or oxygen was found to be
reduced by several orders of magnitude. 

However, the translation of the findings of Ref.~\cite{Oberhummer:2000zj} into 
anthropic constraints on fundamental parameters remains problematic, as the
employed model of the strong force is not readily connected to the fundamental 
theory of  the strong interactions, Quantum Chromodynamics (QCD) and its
fundamental parameters, the light quark masses. 
In this study, we address this question by means of an {\it ab initio}
calculation of the sensitivity of $\varepsilon$ to changes in the light quark
masses and the EM fine structure constant $\alpha_{\rm  em}^{}$.
For this purpose, we carry out large-scale numerical lattice calculations for 
the energies and energy differences relevant to the triple-alpha process
within the framework of chiral Effective Field Theory (EFT). The discretized
(lattice) version of chiral EFT was formulated in Ref.~\cite{Borasoy:2006qn} 
(see Ref.~\cite{Dean_QMC} for a recent review). We have successfully applied 
this novel approach to the spectra and properties
of light
nuclei~\cite{Epelbaum:2009zsa,Epelbaum:2009pd,Epelbaum:2010xt,Epelbaum:2011md}, 
to dilute neutron matter~\cite{Epelbaum:2008vj}, and to the structure of the 
Hoyle state~\cite{Epelbaum:2012qn}. A brief summary of the
results reported here has appeared in Ref.~\cite{Epelbaum:2012iu}.

This paper is organized as follows: In Section~\ref{sec:anatomy}, we briefly 
describe the theoretical framework of the 
present calculation. The quark mass dependence of the nuclear force is
described in Section~\ref{sec:NN} within the framework of 
chiral EFT, in combination with lattice QCD calculations. The analysis 
of the NN system is carried out in Section~\ref{sec:forces}, while 
Section~\ref{sec:AFQMC} deals with the Auxiliary Field Quantum Monte 
Carlo (AFQMC) calculation of the energies of the $^4$He, $^8$Be and $^{12}$C 
ground states as well as the Hoyle state, including the energy differences 
relevant for the triple-alpha process. Section~\ref{sec:HigherOrders} provides 
an estimation of the neglected higher-order effects, while the observed
correlations between the relevant energies and
energy differences are described in Section~\ref{sec:correlations}. 
Finally, in Section~\ref{sec:discussion} the implications of our 
findings for the reaction rate of the triple-alpha process are discussed, and 
Section~\ref{sec:summary} contains a concluding summary.


\section{Theoretical framework} 
\label{sec:anatomy}

As discussed above, the triple-alpha reaction proceeds in two steps. The reaction rate of the first step
$^{4}$He + $^{4}$He $\to$ $^{8}$Be, where the unstable $^{8}$Be resonance is formed, is controlled
by the energy difference $\Delta E_b^{}$,
\begin{equation}
\Delta E_b^{} \equiv E_8^{} - 2 E_4^{}, 
\end{equation}
where we have introduced the notations $E_4^{}$ and $E_8^{}$ for the energies of the ground states
of $^{4}$He and $^{8}$Be, respectively. Also, we shall denote the ground state energy of $^{12}$C by
$E_{12}^{}$, and the energy of the Hoyle state by $E_{12}^\star$. The second step in the triple-alpha process,
$^{8}$Be + $^{4}$He $\to$ $^{12}$C + $\gamma$, depends crucially on the energy difference $\Delta E_h^{}$, 
\begin{equation}
\Delta E_h^{} \equiv E_{12}^\star - E_8^{} - E_4^{},
\end{equation}
such that the reaction rate for fusion of three $\alpha$ particles to $^{12}$C via the 
ground state of $^8$Be and the Hoyle state is given by~\cite{Oberhummer:2000zj}
\begin{equation}
\label{rate}
r_{3 \alpha}^{} = 3^{\frac{3}{2}} N_\alpha^3 
\left( \frac{2 \pi \hbar^2}{|E_4^{}| k_{\rm B}^{} T} \right)^3 
\frac{\Gamma_\gamma^{}}{\hbar} \, \exp \left( -\frac{\varepsilon}{k_{\rm B}^{} T} \right),  
\end{equation}
%
where $N_\alpha^{}$ is the number density
of $\alpha$ particles in a stellar plasma at temperature $T$. The energy 
difference $\varepsilon$ is given by
\begin{equation}
\varepsilon \equiv  \Delta E_b^{} + \Delta E_h^{} = E_{12}^\star - 3 E_4^{},
\end{equation}
which clearly is the dominant control parameter of Eq.~(\ref{rate}) in
comparison with the linear dependence
on the radiative width $\Gamma_\gamma^{}$ of the Hoyle state. The latter can 
therefore be neglected when the impact of small changes in $\varepsilon$ 
on $r_{3 \alpha}^{}$ are considered.

The main question we shall address here is the sensitivity of $\varepsilon$ to 
variations in the light quark masses
and the EM fine structure constant, with the objective of translating the
bounds on $\varepsilon$ found in the stellar
model calculations of Ref.~\cite{Oberhummer:2000zj,Oberhummer:1999ab} to
constraints on these fundamental
parameters. For this purpose, we shall only consider the average light quark
mass $m_q^{} \equiv (m_u^{}+m_d^{})/2$,  as the effects of strong isospin
violation due to $m_u^{} \neq m_d^{}$ are greatly suppressed for the processes 
of relevance to the present work. To this end, we shall calculate how
$E_4^{}$, $E_8^{}$, $E_{12}^{}$ and $E_{12}^\star$ depend on $m_q^{}$ and 
$\alpha_\mathrm{em}$, after which we may use this information to determine
the corresponding dependences of $\Delta E_b^{}$,  $\Delta E_h^{}$ and $\varepsilon$.

Our analysis is carried out within the framework of chiral nuclear EFT, introduced
by Weinberg~\cite{Weinberg:1990rz} as a systematic tool
to explore the consequences of spontaneous and explicit
chiral symmetry breaking of QCD in a rigorous manner. This approach relies on
the most general effective Lagrangian for pions and nucleons
constructed in harmony with the symmetries of QCD. The pions are
identified with the pseudo-Goldstone bosons of the spontaneously broken
chiral symmetry of QCD, which strongly constrains their
interactions. The small (but non-vanishing) pion mass is a result of the 
explicit breaking of chiral symmetry in QCD by the quark masses. 
In particular, one finds $M_\pi^2 \sim (m_u^{}+m_d^{})$, so that any dependence
on the average light quark mass $m_q^{}$ can be translated into a 
corresponding dependence on the pion mass $M_\pi^{}$. 

Chiral nuclear EFT is based on an order-by-order expansion of 
the nuclear potential. In this scheme, two-, three- and four-nucleon forces arise 
naturally, and their observed hierarchy of importance is also explained. 
The nuclear forces have been worked out to high precision and applied successfully
in few-nucleon systems to the binding energies, structure, and reactions 
(see Refs.~\cite{Epelbaum:2008ga,Machleidt:2011zz} for recent reviews).
We have recently developed a discretized version of chiral EFT which allows one to 
compute the correlation function 
\begin{equation}
Z_A^{}(t) = \langle \Psi_A^{} | \exp (-t H) | \Psi_A^{} \rangle, 
\end{equation}
for $A$ nucleons in Euclidean space-time using 
Monte Carlo sampling. Here, $\Psi_A^{}$ denotes the Slater
determinant for $A$ non-interacting nucleons, and $H$ is the nuclear Hamiltonian
calculated in chiral EFT and expressed in terms of the lattice (discretized)
variables. The correlation function $Z_A^{}(t)$ can be efficiently
calculated within the AFQMC
framework, where terms in the lattice action quartic in the nucleon fields
are re-expressed as interactions of a single nucleon with auxiliary fields by 
means of a Hubbard-Stratonovich transformation. Once $Z_A^{} (t)$ has been
calculated, the ground-state energy $E_A^{}$ is  
obtained from the large-$t$ limit of $Z_A^{}$, 
\begin{equation}
E_A^{} = - \lim_{t \to \infty} \frac{d (\ln Z_A^{})}{dt},
\end{equation}
with $t$ the Euclidean time.
We have also developed a multi-channel projection Monte 
Carlo method, which allows us to study excited states by computing the 
correlation matrix for a set of $A$--nucleon states $\Psi_A^i$ with 
appropriately chosen quantum numbers,   
\begin{equation}
Z_A^{ij}(t) = \langle \Psi_A^i | \exp (-t H) | \Psi_A^j \rangle.  
\end{equation}

The AFQMC results reported here correspond to an improved leading-order (LO) action, 
based on the NN amplitude
\begin{eqnarray}
\mathcal{A}_{\rm LO}^{} &= & C_{S=0,I=1}^{} \, f(\vec q) \left( \frac{1}{4} - \frac{1}{4} \,
\vec \sigma_i^{} \cdot \vec \sigma_j^{} \right) \left( \frac{3}{4} + \frac{1}{4} \,
\fet  \tau_i^{} \cdot \fet \tau_j^{} \right) \nonumber \\
&& + \: C_{S=1,I=0}^{} \, f(\vec q) \left( \frac{3}{4} + \frac{1}{4} \,
\vec \sigma_i^{} \cdot \vec \sigma_j^{} \right) \left( \frac{1}{4} - \frac{1}{4} \,
\fet  \tau_i^{} \cdot \fet \tau_j^{} \right) \nonumber \\
&& - \: \tilde g_{\pi N}^2 \, \fet \tau_i^{} \cdot \fet \tau_j^{} \,
\frac{\vec \sigma_i^{} \cdot \vec q \, \vec \sigma_j^{} \cdot \vec q }
{\vec q^2 + M_\pi^2},
\label{LOaction}
\end{eqnarray}
where $\vec \sigma_i^{}$ and $\fet \tau_i^{}$ refer to the Pauli
spin and isospin matrices of nucleon $i$, respectively. The
strength $\tilde g_{\pi N}^{}$ of the one-pion exchange
potential is defined in terms of the nucleon axial-vector coupling $g_A^{}$ 
and the pion decay constant $F_\pi^{}$ as
\begin{equation}
\tilde g_{\pi N}^{} \equiv \frac{g_A^{}}{2 F_\pi^{}},
\end{equation}
while $C_{S=0,I=1}^{}$ and $C_{S=1,I=0}^{}$ are low-energy constants (LECs),
adjusted to reproduce the NN phase shifts in the
$^1S_0^{}$ and $^3S_1^{}$ partial waves respectively. 
The smearing function 
$f(\vec q)$ is chosen to give the (approximately) correct effective ranges
for the two $S$--wave NN channels (see Ref.~\cite{Epelbaum:2010xt} and references
therein for more details, along with a description of the discretized form of the improved LO action).
Corrections of higher order are taken into account in perturbation theory.
In this analysis, we only consider small momentum-independent changes to the short-range interactions.  
These correspond to pointlike contact operators, and it is convenient to express the LECs in terms of the linear combinations 
$C_0^{}$ and $C_I^{}$, 
\begin{eqnarray}
C_0^{} &=& \frac{3}{4} \, C_{S=0,I=1}^{}+\frac{1}{4} \, C_{S=1,I=0}^{}, \label{C0} \\
C_I^{} &=& \frac{1}{4} \, C_{S=0,I=1}^{}-\frac{1}{4} \, C_{S=1,I=0}^{}, \label{CI}
\end{eqnarray}
which couple to the total nucleon density and the isospin density $\fet \tau_i^{} \cdot \fet \tau_j^{}$, respectively.

The expectation value of a given operator $\mathcal O$ is obtained as
\begin{equation}
Z_A^\mathcal O(t) = \langle \Psi_A^{} | \exp (-t H/2) \, \mathcal O 
\, \exp (-t H/2) | \Psi_A^{} \rangle,
\label{expect}
\end{equation}
which accounts for all contributions to the nuclear Hamiltonian up to
next-to-next-to-leading order (N$^2$LO) in the chiral expansion, including the
Coulomb interaction and the three-nucleon forces. 
Our recent AFQMC calculations using this
framework are reported 
in Refs.~\cite{Epelbaum:2009zsa,Epelbaum:2009pd,Epelbaum:2010xt,Epelbaum:2011md,Epelbaum:2012qn}.
In particular, results at N$^2$LO for nuclei with $A = 3,4,6$ and $12$ can be
found in Ref.~\cite{Epelbaum:2010xt}. 
These calculations employ a periodic cubic lattice with a lattice spacing of 
$a = 1.97$~fm and a length of $L = 11.82$~fm. In
the (discretized) Euclidean time direction, we use a step size of $a_t^{} =
1.32$~fm, and perform calculations for propagation times, {\it i.e.} the extent of
the time direction, $L_t^{} \equiv N_t^{} a_t^{}
= 4 \ldots 20$~fm, such that the limit $L_t^{} \to \infty$ is taken by
extrapolation. Given the relatively coarse lattice
spacing employed in our calculations, the two-pion exchange NN potential
that starts contributing at next-to-leading order (NLO) can be well represented
by contact interactions~\cite{Epelbaum:2010xt}.  

We now turn to the $m_q^{}$-dependence or, equivalently, the $M_\pi^{}$-dependence of the
energies $E_i^{}$. We use the notation $E_i^{}$ when referring to either the energies 
$E_4^{}$, $E_8^{}$, $E_{12}^{}$ and $E_{12}^\star$, or to the energy differences 
$\Delta E_b^{}$, $\Delta E_h^{}$ and $\varepsilon$. In this work, we shall restrict the  
values of $M_\pi^{}$ to the vicinity of the physical pion mass, roughly speaking to
$|\delta M_\pi^{}/M_\pi^{}| \leq 10\%$. It is then sufficient to consider the linear 
variation of $E_i^{}$, giving
\begin{equation}
\delta E_i^{} \simeq \frac{\partial E_i^{}}{\partial M_\pi^{}}
\bigg|_{M_\pi^{\rm ph}} \delta M_\pi^{} + 
\left.\frac{\partial E_i^{}}{\partial \alpha_\mathrm{em}^{}} 
\right |_{\alpha_\mathrm{em}^\mathrm{ph}} \delta \alpha_\mathrm{em}^{},
\label{var}
\end{equation}
where we have allowed for independent variations of $M_\pi^{}$ and $\alpha_{\rm em}^{}$.
Our objective is then to compute the partial derivatives in Eq.~(\ref{var}) using AFQMC.
Clearly, such a calculation relies on knowledge of the 
$M_\pi^{}$-dependence of the nuclear Hamiltonian, which is discussed in 
Section~\ref{sec:NN}. We also note that a useful way to express the
sensitivity of a given observable $X$ to a parameter $y$
is given by the dimensionless ``$K$-factors''
\begin{equation} 
K_X^i \equiv \frac{y}{X} \left. \frac{\partial X}{\partial y} \right |_{y^{\rm ph}},
\label{Kfact}
\end{equation}
where we use the superscript $i = \{q, \, \pi, \, \alpha \}$ for the set of observables
$y = \{m_q^{}, \, M_\pi^{}, \, \alpha_{\rm em}^{} \}$. As an example, we can obtain $K_X^q$ 
({\it i.e.} the sensitivity of $X$ to changes in $m_q^{}$) in terms of $K_X^\pi$ by means of
the relation
\begin{equation} 
K_X^q = K_X^\pi K_{M_\pi^{}}^q, 
\end{equation}
where we shall adopt the value $K_{M_\pi^{}}^q =
0.494^{+0.009}_{-0.013}$ from Ref.~\cite{Berengut:2013nh}. 

In addition to shifts in $m_q^{}$, we shall also consider the
effects of shifts in $\alpha_{\rm em}^{}$. The
treatment of the Coulomb interaction in our AFQMC framework is
described in detail in Ref.~\cite{Epelbaum:2010xt}. The main difference between
the continuum and lattice formulations is that the discretized
form of the long-range Coulomb force between two protons
becomes singular if the protons occupy the same lattice site. We therefore employ a
regularized version of the discretized Coulomb interaction, where the
potential energy of two protons on the same lattice site is set to
the continuum value corresponding to a separation of half a lattice spacing. 
The effects of this regularization are compensated for by a derivative-less proton-proton 
contact operator, which also receives contributions from the strong and short-range EM
isospin-breaking effects. The associated coefficient $c_{pp}^{}$ is determined from the 
proton-proton phase shifts. 

The sensitivity of the energies $E_i^{}$ to variations in $\alpha_{\rm em}^{}$
can be obtained by computing the shifts $\Delta E_i^{}(\alpha_{\rm em}^{})$ 
and $\Delta E_i^{}(c_{pp}^{})$. The former is due to the long-range Coulomb 
interaction on the lattice, and the latter arises from the part of the proton-proton contact
operator $\propto c_{pp}^{}$. Specifically, we define
\begin{equation}
\label{defQ}
Q_\mathrm{em}^{} (E_i^{}) \equiv \Delta E_i^{}(c_{pp}^{}) \, x_{pp}^{} 
+ \Delta E_i^{}(\alpha_\mathrm{em}^{}),
\end{equation}
where the coefficient $x_{pp}^{}$ denotes the relative strength of the 
$pp$ contact interaction caused by the regularization of the Coulomb force. 
The determination of $x_{pp}^{}$ will be described in
Section~\ref{sec:AFQMC}.  Given that the energy shifts $\Delta E_i^{}(\alpha_\mathrm{em}^{})$ and 
$\Delta E_i^{}(c_{pp}^{})$ are relatively small, we may approximate 
%
\begin{equation}
\left.\frac{\partial E_i^{}}{\partial \alpha_\mathrm{em}^{}} 
\right |_{\alpha_\mathrm{em}^\mathrm{ph}}
\simeq \frac{Q_\mathrm{em}^{}(E_i^{})}{\alpha_\mathrm{em}^{\rm ph}},
\end{equation}
with $\alpha_{\rm em}^{\rm ph} \simeq 1/137$. While we shall mainly study the
explicit dependence on $\alpha_{\rm em}^{}$ induced by the Coulomb interaction, 
it is worth noting that $K_{X}^\alpha$ may receive additional
contributions from the corresponding shifts in the effective
had\-ro\-nic Lagrangian, for example from the EM shift of $M_\pi^{}$. 
Schematically, this may be expressed as
\begin{equation}
K_{X}^\alpha = \frac{Q_\mathrm{\rm em}^{}(X)}{X} + \frac{\Delta M_\pi^\mathrm{em}}{X}
\left.\frac{\partial X}{\partial M_\pi^{}} \right
|_{M_\pi^\mathrm{ph}} + \ldots,
\end{equation}
where the sizes of such additional contributions to the EM $K$-factors are 
$\Delta M_\pi^{\rm em}/M_\pi^{} \simeq 4\%$ compared to the dominant strong
contributions, and are thus not considered in the present analysis.  


\section{Pion mass dependence of the nuclear Hamiltonian} 
\label{sec:NN}

The $M_\pi^{}$-dependence of the energies $E_i^{}$ is generated by the
$M_\pi^{}$-dependence of the nucleon mass $m_N^{}$ in the kinetic energy term in
the nuclear Hamiltonian, as well as by the $M_\pi^{}$-dependence of the nuclear potentials.
In the present analysis, we will not take into account the sources of $M_\pi^{}$-dependence 
generated by contributions beyond $\mathcal{A}_{\rm LO}^{}$ in
Eq.~(\ref{LOaction}). Instead, we shall
estimate the neglected higher-order contributions to the energy shifts in 
Section~\ref{sec:HigherOrders}.  On the other hand, in order to obtain the
most accurate description possible of the $M_\pi^{}$-dependence of
$\mathcal{A}_{\rm LO}^{}$ and to ensure model independence, we shall go beyond
the strict chiral expansion of the terms entering Eq.~(\ref{LOaction}), and make use 
of the available lattice QCD data whenever possible.   

For the $M_\pi^{}$-dependence of the nucleon mass, we define the 
quantity
\begin{equation} 
x_1^{} \equiv \left. \frac{\partial m_N^{}}{\partial M_\pi^{}} 
\right|_{M_\pi^\mathrm{ph}},
\end{equation}
which has been analyzed extensively in the
literature by combining Chiral Perturbation Theory (ChPT) with
lattice QCD data, see {\it e.g.} Ref.~\cite{Mnucl,Bernard:2007zu}.  
At $\mathcal{O}(p^2)$ in ChPT, $m_N^{}(M_\pi^{})$ is given by
\begin{equation}
m_N^{} = m_0^{} - 4c_1^{} M_\pi^2 + \mathcal{O}(p^3),
\label{mN}
\end{equation}
where $m_0^{}$ denotes the value of $m_N^{}$ in the chiral limit. 
The value $x_1^{} \simeq 0.57$ was obtained in Ref.~\cite{Berengut:2013nh},
which corresponds to the $\mathcal{O}(p^3)$ heavy-baryon~(HB) ChPT result with  
$c_1^{} = -0.81$~GeV$^{-1}$. Alternatively, one may determine 
$x_1^{}$ from the pion-nucleon sigma term 
$\sigma_{\pi N}^{} \equiv M_\pi^2 \, \partial m_N^{} / \partial M_\pi^2$ 
by means of the Feynman-Hellmann theorem. 
From the results of Ref.~\cite{Frink:2005ru}, one finds
\begin{equation}
\sigma_{\pi N}^{} = 44.9^{+1.8}_{-5.4}~\mathrm{MeV} \quad \to \quad
x_1^{} = 0.66^{+0.02}_{-0.08},
\end{equation}
while those of Ref.~\cite{Alarcon:2011zs} yield
\begin{equation}
\sigma_{\pi N}^{} = 59 \pm 7~\mathrm{MeV} \quad \to \quad
x_1^{} = 0.87 \pm 0.10,
\end{equation}
and we also note that
$x_1^{} \simeq 0.73$ has been obtained by Procura {\it et al.}
from fits of a modified $\mathcal{O}(p^4)$ ChPT formula to lattice 
QCD data~\cite{Mnucl}. In our analysis, we adopt the conservative estimate 
\begin{equation}
\label{valx1}
x_1^{} = 0.57 \ldots 0.97,   
\end{equation}
based on the variation in the results quoted above.

We now turn to the $M_\pi^{}$-dependence of the nuclear force.  The most 
obvious source of $M_\pi^{}$-dependence is the static pion propagator in 
Eq.~(\ref{LOaction}). In addition to this  explicit dependence on $M_\pi^{}$, 
we also take into account the implicit $M_\pi^{}$-dependence
of the coupling constant $\tilde g_{\pi N}^{}$ of the one-pion exchange
(OPE) potential by defining
\begin{equation}
x_2^{} \equiv
\left.\frac{\partial \tilde g_{\pi N}^{}}{\partial M_\pi^{}}\right|_{M_\pi^\mathrm{ph}}
 = \frac{1}{2 F_\pi^{}} 
\left.\frac{\partial g_A^{}}{\partial M_\pi^{}}\right|_{M_\pi^\mathrm{ph}}
- \frac{g_A^{}}{2 F_\pi^2} 
\left.\frac{\partial F_\pi^{}}{\partial M_\pi^{}}\right|_{M_\pi^\mathrm{ph}},
\label{x2val}
\end{equation}
where both contributions to $x_2^{}$ have been studied extensively
by means of ChPT and lattice QCD 
(see Refs.~\cite{Berengut:2013nh,Bernard:2007zu,Colangelo} and references
therein). As for $m_N^{}$ in Eq.~(\ref{mN}), we obtain
\begin{equation}
F_\pi^{} = F \left(1 + 
\frac{M_\pi^2}{16\pi^2 F^2} \bar l_4^{} + \mathcal{O}(M_\pi^4) \right),
\end{equation}
where we use $\bar l_4^{} \simeq 4.3$ from Ref.~\cite{Gasser:1983yg},
which is consistent with modern lattice determinations, see e.g.
Ref.~\cite{Baron:2010bv},  $\bar l_4=4.67(3)(10)$,
and $F \simeq 86.2$~MeV is the value of $F_\pi^{}$ in the chiral limit. 
As an alternative to the sub-leading order ChPT result, one may use the 
determination of $K_{F_\pi^{}}^q$ reported   
in Ref.~\cite{Berengut:2013nh}, $K_{F_\pi^{}}^q = 0.048 \pm 0.012$, which
is based on a combined analysis in ChPT and lattice QCD. This gives
\begin{equation}
\label{dfpidmpi}
\left.\frac{\partial F_\pi^{}}{\partial M_\pi^{}}\right|_{M_\pi^\mathrm{ph}}  =
\frac{F_\pi^{}}{M_\pi^{}} \, \frac{K_{F_\pi^{}}^q}{K_{M_\pi^{}}^q} \simeq 0.066, 
\end{equation}
using the central value $K_{M_\pi^{}}^q = 0.494$, as discussed in Section~\ref{sec:anatomy}.

While the chiral expansion of $F_\pi^{}$ shows good convergence, the equivalent
expression for $g_A^{}$ is known to converge much slower.
In particular, the $\mathcal{O}(p^3)$ HB ChPT result for $g_A^{}$ shows a very strong 
$M_\pi^{}$-dependence near the physical
point (see Refs.~\cite{Berengut:2013nh,Bernard:2006te} and references
therein). On the other hand, lattice QCD calculations indicate that 
$g_A^{}$ as a function of $M_\pi^{}$ is remarkably flat. In principle, 
such flat behavior can be accommodated at the two-loop level in 
ChPT~\cite{Berengut:2013nh}.  By assuming that $g_A^{}$ is constant 
in the vicinity of the physical point, the value $x_2^{} \simeq -0.049$ 
in lattice units (l.u.) is obtained from Eq.~(\ref{x2val}).
This should be compared with the central value reported in 
Ref.~\cite{Berengut:2013nh}, $x_2^{} =
-0.024$~l.u., which is obtained from the incomplete $\mathcal{O}(p^4)$~HB result for 
$g_A^{}$, constrained by the available lattice QCD result at $M_\pi^{} \simeq 350$~MeV. 
The relatively large uncertainty in the chiral extrapolation of
$g_A^{}$ has a significant impact on the allowed values of $x_2^{}$. The
largest source of uncertainty in Ref.~\cite{Berengut:2013nh} is the poorly known
low-energy constant $\bar d_{16}^{}$. The present empirical constraints from
the reaction $\pi N \to \pi \pi N$ yield a relatively large range of  
$\bar d_{16}^{} = -0.91 \ldots -\!2.61~\mathrm{GeV}^{-2}$, which in turn gives
\begin{equation}
\label{valx2}
x_2^{} = -0.056 \ldots 0.008 \:\: \mbox{l.u.},
\end{equation}
for the range of uncertainty in $x_2^{}$, which we shall adopt 
in the present analysis.

The short-range part of the nuclear force also depends on $M_\pi^{}$.
This dependence is more difficult to control within chiral EFT,
see Ref.~\cite{Berengut:2013nh} for an extended discussion. Since we aim at
a model-independent determination of the $M_\pi^{}$-dependence of the nuclear
energies $E_i^{}$, we refrain from a chiral expansion of the short-range
part of the nuclear force. Rather, we shall \emph{parameterize} the
$M_\pi^{}$-dependence of the LO contact interactions, {\it i.e.} of the
coefficients $C_0^{}$ and $C_I^{}$ in Eq.~(\ref{LOaction}). This can be 
performed in terms of the slope  of the inverse NN $S$--wave scattering 
lengths $a_{s}^{-1}$ and
$a_{t}^{-1}$, 
\begin{equation}
\label{defA}
\bar A_{s}^{} \equiv \frac{\partial a_{s}^{-1}}{\partial M_\pi^{}}
\bigg|_{M_\pi^{\rm ph}}, 
\quad
\bar A_{t}^{} \equiv \frac{\partial a_{t}^{-1}}{\partial M_\pi^{}}
\bigg|_{M_\pi^{\rm ph}},
\end{equation}
where we have introduced the subscripts $s$ and $t$ for the 
spin--$0$ ($^1S_0^{}$) and spin--$1$ ($^3S_1^{}$) NN partial waves. 
For the purpose of our analysis, $\bar A_{s}^{}$ and $\bar A_{t}^{}$ are
regarded as input parameters, and we shall express all our results in terms of these.
We shall return to the determination of $\bar A_{s}^{}$ and $\bar A_{t}^{}$ 
in Section~\ref{sec:discussion}.

Given the sources of $M_\pi^{}$-dependence discussed so far, 
we may express the dependence of the energies $E_i^{}$ on $M_\pi^{}$ as
\begin{equation}
E_i^{} = E_i^{}(\tilde M_\pi^{},
m_N^{}(M_\pi^{}), \tilde g_{\pi N}^{}(M_\pi^{}), 
C_0^{}(M_\pi^{}), C_I^{}(M_\pi^{})),
\label{Eeq}
\end{equation}
where $\tilde M_\pi^{}$ refers to the explicit $M_\pi^{}$-dependence from the pion propagator in
the OPE contribution. In order to assess the sensitivity of the triple-alpha
process (and of the various energy levels involved in that process) to shifts
in $M_\pi^{}$, we will compute quantities of the form 
$\partial E_i^{}/\partial M_\pi^{}$ at the physical point. 
Given Eq.~(\ref{Eeq}), we find
\begin{eqnarray}
\left. \frac{\partial E_i^{}}{\partial M_\pi^{}} \right|_{M_\pi^\mathrm{ph}} &= &
\left. \frac{\partial E_i^{}}{\partial \tilde M_\pi^{}} \right|_{M_\pi^\mathrm{ph}}
+ x_1^{} \left. \frac{\partial E_i^{}}{\partial m_N^{}} \right|_{m_N^\mathrm{ph}}
+ x_2^{} \left. \frac{\partial E_i^{}}{\partial \tilde g_{\pi N}^{}} \right|_{\tilde g_{\pi N}^\mathrm{ph}} 
\nonumber \\
& +& x_3^{} \left. \frac{\partial E_i^{}}{\partial C_0^{}}
\right|_{C_0^\mathrm{ph}}
+ x_4^{} \left. \frac{\partial E_i^{}}{\partial C_I^{}} \right|_{C_I^\mathrm{ph}},
\label{Eeq2}
\end{eqnarray}
where we have introduced the definitions
\begin{align}
& x_3^{} \equiv \left. \frac{\partial C_0^{}}{\partial M_\pi^{}} \right|_{M_\pi^\mathrm{ph}}, \quad
x_4^{} \equiv \left. \frac{\partial C_I^{}}{\partial M_\pi^{}} \right|_{M_\pi^\mathrm{ph}},
\label{xy}
\end{align}
for the short-range components of the LO amplitude. Our method for re-expressing the scheme-dependent
parameters $x_3^{}$ and $x_4^{}$ in terms of $\bar A_{s}^{}$ and $\bar A_{t}^{}$ is 
explained in Section~\ref{sec:forces}. 
The AFQMC calculation of the partial derivatives in Eq.~(\ref{Eeq2}) 
is detailed in Section~\ref{sec:AFQMC}.  
 

\section{Short-range contributions to the nuclear force} 
\label{sec:forces}

With the exception of the OPE contribution, much of the $M_\pi^{}$-dependence 
of the nuclear Hamiltonian is implicit and thus controlled by the coefficients 
$x_i^{}$, which describe how $m_N^{}, \tilde g_{\pi N}^{}, C_0^{}$ and  
$C_I^{}$ depend on $M_\pi^{}$. As discussed in Section~\ref{sec:NN}, $x_1^{}$
and $x_2^{}$ are fairly well constrained  by ChPT calculations in combination
with lattice QCD data, and provide external input to our analysis according to
Eqs.~(\ref{valx1}) and~(\ref{valx2}). 
Contrary to $x_1^{}$ and $x_2^{}$, 
the coefficients $x_3^{}$ and $x_4^{}$ are strongly scheme-dependent. 
We shall therefore express these in terms of the scheme-independent quantities 
$\bar A_s^{}$ and $\bar A_t^{}$. This
substitution is achieved by an analysis of the two-nucleon scattering problem 
on a periodic lattice. Once 
the dependence of the $S$--wave scattering lengths in the spin-singlet
and triplet channels on $M_\pi^{}$ is known, $x_3^{}$ and $x_4^{}$ can be 
straightforwardly obtained.

We use the finite volume formula due to L\"uscher~\cite{Luscher:1986pf,Luscher:1990ux} which relates the energy 
levels of a two-body system in a cubic periodic volume of length $L$ to the scattering phase shifts.  For the $S$-wave case, we have
\begin{equation}
p \cot \delta = \frac{1}{\pi L} S(\eta) \approx -\frac{1}{a}, 
\quad\quad
\eta \equiv \left(\frac{pL}{2\pi}\right)^2,
\label{pcotdelta}
\end{equation}
where the three-dimensional zeta function $S(\eta)$ is given by
\begin{equation}
S(\eta)=\lim_{\Lambda\rightarrow\infty}\left[  \sum_{\vec{n}}\frac
{\theta(\Lambda^{2}-\vec{n}^{2})}{\vec{n}^{2}-\eta}-4\pi\Lambda\right].
\end{equation}
For $\left\vert \eta\right\vert <1$, it is convenient to expand $S(\eta)$ in powers 
of $\eta$ as shown in Ref.~\cite{Lee:2007a}. We now differentiate Eq.~(\ref{pcotdelta}) with respect to $M_\pi^{}$, which yields 
\begin{equation}
\frac{\partial a^{-1}_{}}{\partial M_\pi^{}}  
= - \frac{1}{\pi L} S'(\eta) \frac{\partial \eta}{\partial M_\pi^{}}.
\end{equation}
If we denote the two-nucleon energy in the singlet channel by $E_s^{}$ and that in the triplet
channel by $E_t^{}$, we find
\begin{equation}
\frac{\partial \eta_{s,t}^{}}{\partial M_\pi^{}} = 
\left(\frac{L}{2\pi}\right)^2_{}
\left( E_{s,t}^{} \frac{\partial m_N^{}}{\partial M_\pi^{}} + 
m_N^{} \frac{\partial E_{s,t}^{}}{\partial M_\pi^{}} \right), 
\end{equation}
and by introducing the notation of Eq.~(\ref{defA}), we
obtain the relations
\begin{eqnarray}
- \zeta_s^{-1} \, \bar A_s^{} &=& 
\left.\frac{\partial E_s^{}}{\partial \tilde M_\pi^{}} \right|_{M_\pi^\mathrm{ph}}
+ x_1^{} \left( \frac{E_s^{}}{m_N^{}} + \left.\frac{\partial E_s^{}}{\partial
      m_N^{}} \right|_{m_N^\mathrm{ph}} \right) \nonumber \\
&& + \: x_2^{} \left.\frac{\partial E_s^{}}{\partial \tilde g_{\pi N}^{}} 
\right|_{\tilde g_{\pi N}^\mathrm{ph}}
+ (x_3^{} + x_4^{}) \, q_s^{}, 
\label{equationAs} \\
- \zeta_t^{-1} \, \bar A_t^{} &=& 
\left.\frac{\partial E_t^{}}{\partial \tilde M_\pi^{}} \right|_{M_\pi^\mathrm{ph}}
+ x_1^{} \left( \frac{E_t^{}}{m_N^{}} + \left.\frac{\partial E_t^{}}{\partial
      m_N^{}} \right|_{m_N^\mathrm{ph}} \right) \nonumber \\
&& + \: x_2^{} \left.\frac{\partial E_t^{}}{\partial \tilde g_{\pi N}^{}} 
\right|_{\tilde g_{\pi N}^\mathrm{ph}} + (x_3^{} - 3  x_4^{}) \, q_t^{}, 
\label{equationAt}
\end{eqnarray}
where we have defined
\begin{equation}
\zeta_{s,t}^{} \equiv \frac{m_N^{}L}{4\pi^3} S'(\eta_{s,t}^{}), \quad \quad 
q_{s,t}^{} \equiv \left.\frac{\partial E_{s,t}^{}}{\partial C_0^{}} 
\right|_{C_0^\mathrm{ph}}~.
\end{equation}

\begin{table}[t]
\centering{
\caption{
LO energies (in MeV) of the spin-singlet ($E_s^{}$) and spin-triplet ($E_t^{}$) 
two-nucleon states used in the L\"uscher analysis, along with the required 
partial derivatives. The results were obtained by numerical solution 
of the Schr\"odinger equation in a cubic box 
of size $N = 24$ (second column) and $N=32$ (third column). 
$E_d^{}$ denotes the energy of the deuteron.
Quantities labeled ``[l.u.]'' are given in units of the inverse 
(spatial) lattice spacing.  All derivatives are evaluated at the physical point.
\label{tab_scatt_24}}
\vspace{.3cm}
\begin{tabular}{|c|c|c|}
\hline 
& $L = 47.36~\mathrm{fm}$ & $L = 63.14~\mathrm{fm}$   \\
\hline
$E_s^{}$(LO) [MeV] & $-0.0440956$ &  $-0.0218593$\\
$E_t^{}$(LO) [MeV] & $0.0369820$  &  $0.0142463$  \\ 
$E_d^{}$(LO) [MeV] & $-2.2244401$ & $-2.2443719$ \\
$\partial E_s^{} / \partial \tilde M_\pi^{}$ & $-2.0261727 \! \times \!
10^{-4}$ &  $-1.1927443 \! \times \! 10^{-4}$ \\
$\partial E_t^{} / \partial \tilde M_\pi^{}$ & $-5.5518808 \! \times \!
10^{-5}$ & $-1.9921726 \! \times \! 10^{-5}$ \\
$\partial E_s^{} / \partial m_N^{}$ & $-2.0081283 \! \times \! 10^{-4}$ & $-1.2224536 \! \times \! 10^{-4}$ \\
$\partial E_t^{} / \partial m_N^{}$ &  $-1.6013626 \! \times \! 10^{-4}$ & $-5.8480992 \! \times \! 10^{-5}$ \\ 
$\partial E_s^{} / \partial \tilde g_{\pi N}^{}$[l.u.] & $9.9946261
\! \times \! 10^{-4}$ &  $5.8797858 \! \times \! 10^{-4}$ \\
$\partial E_t^{} / \partial \tilde g_{\pi N}^{}$[l.u.] & $3.5402207
\! \times \! 10^{-4}$ & $1.2691192 \! \times \! 10^{-4}$ \\ 
$q_s^{}$ [l.u.] & $0.00379650$ &  $0.00223108$ \\
$q_t^{}$ [l.u.]  & $0.00165886$ & $5.9467427 \! \times \! 10^{-4}$ \\
\hline
\end{tabular}
}
\end{table} 

Our results for the energies $E_{s,t}^{}$ and the corresponding partial derivatives, including the 
factors $q_{s,t}^{}$, are summarized in Table~\ref{tab_scatt_24}. These are computed by
exact numerical solution of the two-nucleon problem on a spatial lattice. As the objective is 
to take the box size $N$ (with $L = Na$) large enough to make finite volume effects negligible, 
two different box sizes ($N = 24$ and $N = 32$) have been considered in order to determine the 
magnitude of residual finite volume effects on the L\"uscher analysis.
Given Eqs.~(\ref{equationAs}) and~(\ref{equationAt}) and the results in 
Table~\ref{tab_scatt_24}, we are in the position to compute $x_3^{}$ and
$x_4^{}$ for use in Eq.~(\ref{Eeq2}). However, it is also instructive
to eliminate $x_3^{}$ and $x_4^{}$ analytically, which provides an alternative 
to the parametrization of Eq.~(\ref{Eeq2}). This is particularly useful when
we express our final results in terms of  $\bar A_s^{}$ and $\bar A_t^{}$, as 
the current knowledge of these parameters contains sizable uncertainties. 
Elimination of $x_3^{}$ and $x_4^{}$ in favor of $\bar A_s^{}$ and $\bar A_t^{}$ gives
\begin{align}
\left. \frac{\partial E_i^{}}{\partial M_\pi^{}} \right|_{M_\pi^\mathrm{ph}} \equiv &
-\frac{Q_s^\mathrm{MC}}{\zeta_s^{}} \, \bar A_s^{}
-\frac{Q_t^\mathrm{MC}}{\zeta_t^{}} \, \bar A_t^{}
- Q_s^\mathrm{MC} R_s^{}(x_1^{},x_2^{}) \nonumber \\
& - Q_t^\mathrm{MC} R_t^{}(x_1^{},x_2^{}) + R_\mathrm{MC}^{}(x_1^{},x_2^{}),
\label{QR}
\end{align}
which is equivalent to Eq.~(\ref{Eeq2}). Here, the notation ``MC'' indicates
which quantities incorporate information from the AFQMC framework. In this 
parametrization, the individual terms may be obtained
using the numbers in Table~\ref{tab_scatt_24}, together with the AFQMC results 
of Section~\ref{sec:AFQMC}. In Eq.~(\ref{QR}), we have introduced the quantities
\begin{align}
Q_s^\mathrm{MC} & \equiv 
\frac{3}{4q_s^{}} \left. \frac{\partial E_i^{}}{\partial C_0^{}} \right|_{C_0^\mathrm{ph}}
+ \frac{1}{4q_s^{}} \left. \frac{\partial E_i^{}}{\partial C_I^{}} \right|_{C_I^\mathrm{ph}}, \\
Q_t^\mathrm{MC} & \equiv 
\frac{1}{4q_t^{}} \left. \frac{\partial E_i^{}}{\partial C_0^{}} \right|_{C_I^\mathrm{ph}}
- \frac{1}{4q_t^{}} \left. \frac{\partial E_i^{}}{\partial C_I^{}} \right|_{C_I^\mathrm{ph}},
\end{align}
for which the error is given entirely by the statistical uncertainty of the AFQMC calculation.
We also have
\begin{align}
R_s^{}(x_1^{},x_2^{}) \equiv & 
\left.\frac{\partial E_s^{}}{\partial \tilde M_\pi^{}}\right|_{M_\pi^\mathrm{ph}}
+x_1^{} 
\left( \frac{E_s^{}}{m_N^{}} + \left.\frac{\partial E_s^{}}{\partial m_N^{}} 
\right|_{m_N^\mathrm{ph}} \right) 
\nonumber \\
& + x_2^{} 
\left.\frac{\partial E_s^{}}{\partial \tilde g_{\pi N}^{}} \right|_{\tilde g_{\pi N}^\mathrm{ph}}, \\
R_t^{}(x_1^{},x_2^{}) \equiv & 
\left.\frac{\partial E_t^{}}{\partial \tilde M_\pi^{}}\right|_{M_\pi^\mathrm{ph}}
+x_1^{} 
\left( \frac{E_t^{}}{m_N^{}} + \left.\frac{\partial E_t^{}}{\partial m_N^{}} \right|_{m_N^\mathrm{ph}} \right) 
\nonumber \\
& + x_2^{} \left.\frac{\partial E_t^{}}{\partial \tilde g_{\pi N}^{}} \right|_{\tilde g_{\pi N}^\mathrm{ph}},
\end{align}
where, in contrast, the dominant sources of uncertainty come from $x_1^{}$ and $x_2^{}$. 
Finally, we have
\begin{equation}
R_\mathrm{MC}^{}(x_1^{},x_2^{}) \equiv 
\left. \frac{\partial E_i^{}}{\partial \tilde M_\pi^{}} \right|_{M_\pi^\mathrm{ph}}
+ x_1^{} \left. \frac{\partial E_i^{}}{\partial m_N^{}} \right|_{m_N^\mathrm{ph}}
+ x_2^{} \left. \frac{\partial E_i^{}}{\partial \tilde g_{\pi N}^{}} \right|_{\tilde g_{\pi N}^\mathrm{ph}},
\label{RMC}
\end{equation}
which combines the AFQMC results for the OPE and kinetic energy
contributions. The error of $R_\mathrm{MC}^{}$
receives contributions from the statistical AFQMC error as
well as from $x_1^{}$ and $x_2^{}$. 

While the most convenient way to obtain our final results is by means of 
Eq.~(\ref{QR}), we may also use the values given in Table~\ref{tab_scatt_24} 
to solve Eqs.~(\ref{equationAs}) and~(\ref{equationAt})  for $x_3^{}$ and
$x_4^{}$, and to express these as functions of $\bar A_{s,t}^{}$ and
$x_{1,2}^{}$.  This also allows us to illustrate the size of the finite volume 
effects in the L\"uscher analysis. We obtain the relations
\begin{align}
x_3^{} & = 4.8394 \times 10^{-2}
+ 6.7146 \times 10^{-2}  \, x_1^{} \\ 
& \quad - 0.25080 \, x_2^{} 
- 0.37540 \, \bar A_s^{} - 0.20377 \, \bar A_t^{}, \nonumber \\
x_4^{} & = 4.9754 \times 10^{-3}
- 1.8813 \times 10^{-3} \, x_1^{} \\ 
& \quad - 1.2462 \times 10^{-2} \, x_2^{}
- 0.12513 \, \bar A_s^{} + 0.20377 \, \bar A_t^{}, \nonumber 
\end{align}
for the smaller $N = 24$ lattice ($L=47.36$~fm), and 
\begin{align}
x_3^{} & = 4.8470 \times 10^{-2}
+ 6.7127 \times 10^{-2} \, x_1^{} \\ 
& \quad - 0.25101 \, x_2^{} 
- 0.37652 \, \bar A_s^{} - 0.20467 \, \bar A_t^{}, \nonumber \\
x_4^{} & = 4.9901 \times 10^{-3}
- 1.8998 \times 10^{-3} \, x_1^{} \\
& \quad - 1.2532 \times 10^{-2} \, x_2^{}
- 0.12551 \, \bar A_s^{} + 0.20467 \, \bar A_t^{}, \nonumber
\label{elim_x3x4}
\end{align}
for the larger $N = 32$ lattice ($L=63.14$~fm). In the above equations, the
dimensionful quantities $x_2^{}$,  $x_3^{}$ and $x_4^{}$
should be taken in units of the corresponding powers of the inverse
lattice spacing. We note that the results for $N = 24$ and $N = 32$ are practically
indistinguishable. In Section~\ref{sec:AFQMC}, we shall make use of the 
$N = 32$ lattice when presenting our final results.


\section{Auxiliary Field Quantum Monte Carlo results} 
\label{sec:AFQMC} 

We now turn to the AFQMC calculation of the shifts in the $E_i^{}$. Our 
Monte Carlo simulations are  performed for a single value of $M_\pi^{}$, 
equal to the neutral pion mass, with isospin symmetry breaking treated as 
a perturbation. The partial derivatives $\partial E_i^{} / \partial C_0^{}$ and
$\partial E_i^{} / \partial C_I^{}$ are obtained by computing the matrix
elements of the associated operators according to Eq.~(\ref{expect}). On the
other hand, the partial derivatives $\partial E_i^{} / \partial \tilde
M_\pi^{}$ with respect to the pion mass in the OPE term are computed by
evaluating in perturbation theory the energy shift $\Delta E_i^{}$ induced by 
the substitution $H(\tilde M_\pi^{}) \to H(\tilde M_\pi^{} + \Delta \tilde
M_\pi^{})$ in the nuclear Hamiltonian. Here, the 
masses of both the neutral and charged pions in the OPE term have been shifted 
by $\Delta M_\pi^{} = 4.59$~MeV, which equals 
the empirical mass difference between the neutral and charged pions. 
The corresponding partial derivatives $\partial E_i^{} / \partial \tilde M_\pi^{}$ that 
enter Eqs.~(\ref{Eeq2}) and~(\ref{RMC}) are then given by 
\begin{equation}
\left.
\frac{\partial E_i^{}}{\partial \tilde M_\pi^{}} 
\right |_{M_\pi^{\rm ph}} \simeq \frac{\Delta E_i^{} 
(\Delta \tilde M_\pi^{})}{\Delta M_\pi^{}},
\end{equation}
which we find accurate to within the statistical error of the AFQMC
calculation. We shall also briefly consider the closely related energy shift
$\Delta E_i^{} (\Delta \tilde M_\pi^\mathrm{IB})$, given by the substitution 
$H(\tilde M_\pi^{}) \to H(\tilde M_\pi^{} + \Delta \tilde M_\pi^\mathrm{IB})$ in the
nuclear Hamiltonian. In this case, only the masses of the charged pions are
shifted to their physical value. Finally, the
partial derivatives $\partial E_i^{} / \partial \tilde g_{\pi N}^{}$ and 
$\partial E_i^{} / \partial \tilde m_N^{}$ are obtained as a finite difference,
by defining the quantities $\tilde g_{\pi N}^{} \pm \Delta \tilde g_{\pi
  N}^{}$ and $m_N^{} \pm \Delta m_N^{}$, followed by computation of the resulting
shifts of the $E_i^{}$ in perturbation theory.

\begin{table}[t]
\centering{
\caption{Validation of the extrapolation $N_t^{} \to \infty$ to infinite
  Euclidean time. The LO deuteron energy $E_d^{}$ and the 
corresponding energy shifts and derivatives (at the physical point)
are computed using AFQMC and extrapolated $N_t^{} \to \infty$ (second column) 
and compared with the values obtained from the numerical solution of the 
Schr\"odinger equation (third column). The appropriate units are given
for each quantity, with ``[l.u.]'' indicating units of the inverse 
(spatial) lattice spacing. Parentheses indicate one-standard-deviation
errors. 
\label{tab_deut}}
\vspace{.3cm}
\begin{tabular}{|l| r | r|}
\hline
& \multicolumn{1}{c|}{$^2$H (MC+ex)}
& \multicolumn{1}{c|}{$^2$H (exact)} 
\\ \hline
$E_d^{} (\mathrm{LO})$ [MeV]
& $-9.070(12)$ & $-9.078$ \\
$\Delta E_d^{} (\Delta \tilde M_\pi^{})$ [MeV]
& $-0.003548(12)$ & $-0.003569$ \\  
$\Delta E_d^{} (\Delta \tilde M_\pi^\mathrm{IB})$ [MeV]
& $-0.002372(8)$ & $-0.002379$ \\ 
$\partial E_d^{} / \partial m_N^{}$%
& $-0.00382(2)$ & $-0.003809$ \\
$\partial E_d^{} / \partial \tilde g_{\pi N}^{}$ [l.u.]
& $0.01024(11)$ & $0.01017$ \\
$\partial E_d^{} / \partial C_0^{}$ [l.u.]
& $0.13897(15)$ & $0.138867$ \\
$\partial E_d^{} / \partial C_I^{}$ [l.u.] 
& $-0.4171(4)$ & $-0.41660$\\
\hline
\end{tabular}
}
\end{table}

All AFQMC results presented here have been
extrapolated to infinite Euclidean time ($N_t^{} = \infty$). Such an
extrapolation is necessary, as increasing
the number of Euclidean time steps beyond $N_t^{} = 14$ for $^4$He and $N_t^{} = 12$
for the heavier nuclei becomes impractical due to the worsening sign problem.
An accurate extrapolation is necessary for reliable conclusions. As an
example, on an $N = 6$ lattice the 
calculated $^4$He binding energy at $N_t^{} = 14$ still deviates from the
extrapolated value  at
the $\sim 10$\% level. 
The extrapolation of the LO energies $E_i^{}$ is performed using the trial function
\begin{equation}
E_i^{}(N_t^{}) = E_i^{}(\infty) + c_{E,i}^{}
\exp\left(-\frac{N_t^{}}{\tau_i^{}}\right),
\label{extr1}
\end{equation}
and all matrix elements computed in perturbation theory (energy shifts and
partial derivatives, collectively labeled $X_i^{}$) are extrapolated with
\begin{equation}
X_i^{}(N_t^{}) = X_i^{}(\infty) + c_{X,i}^{}
\exp\left(-\frac{N_t^{}}{2\tau_i^{}}\right),
\label{extr2}
\end{equation}
using the exponent $\tau_i^{}$ from Eq.~(\ref{extr1}). The extrapolation is performed by means of a 
simultaneous chi-square minimization for all quantities, such
that the fitted parameters are $E_i^{}(\infty), X_i^{}(\infty), c_{E,i}^{},
c_{X,i}^{}$ and $\tau_i^{}$. The AFQMC data for each nucleus is fitted with a distinct correlation length $\tau_i^{}$.
In general, we observe that the rate of convergence with $N_t^{}$ is larger for the $^{12}$C ground and Hoyle states 
than for $^{4}$He and $^{8}$Be.

We have also investigated sources of systematical error that arise from the restriction of the extrapolation to
a single exponential $\tau_i^{}$, which is taken to be common for all matrix elements in a given channel. We find that the stability of the 
single-exponential extrapolations for $^4$He and $^8$Be
requires that the data for the matrix elements be excluded for $N_t^{} < 6$, and 
that $\tau_i^{}$ depends significantly on the choice of trial wave function. We therefore conclude that the
single-exponential {\it ansatz}  is more reliable for $^{12}$C. A more accurate extrapolation should allow for multiple
exponentials, which may affect each matrix element in a given nuclear channel to a varying extent. In order to reliably perform
such an analysis, substantially more AFQMC data is required, including data for multiple trial wave functions. At present, the
uncertainties in $\bar A_{s,t}^{}$ and $x_{1,2}^{}$ clearly outweigh the additional systematical error introduced by the restriction
of the extrapolation to a single exponential $\tau_i^{}$ for each nuclear channel.

\begin{figure}[t]
\centering
\includegraphics[width = \columnwidth]{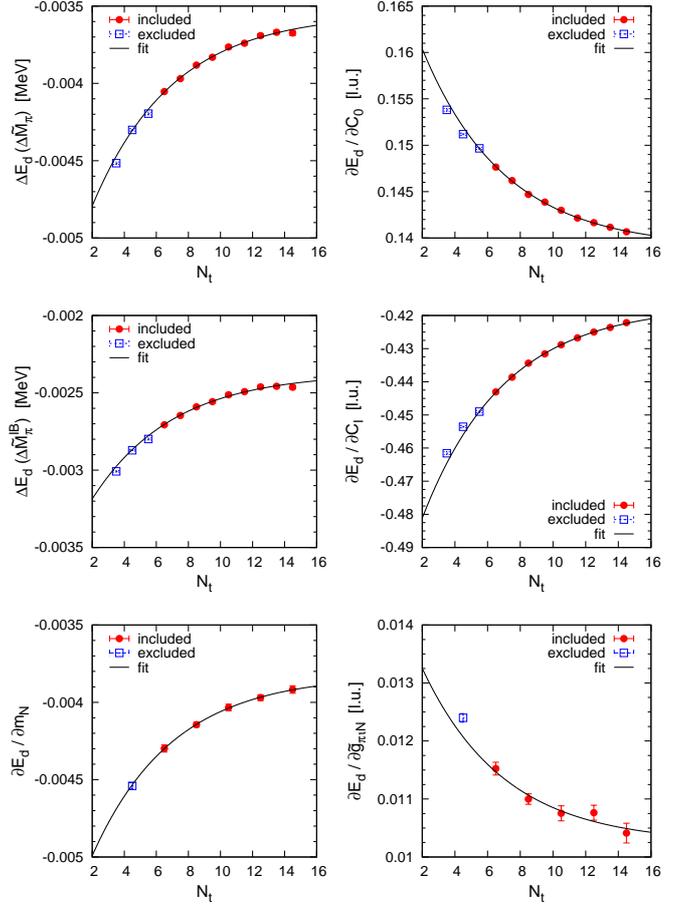}
\caption{AFQMC calculation of the deuteron, as a function of Euclidean 
time $N_t^{}$. Results after the extrapolation $N_t^{} \to \infty$ are given 
in Table~\ref{tab_deut}. The results for $E_d^{}$(LO) have been omitted, 
as they show no appreciable dependence on $N_t^{}$, and can thus be 
trivially extrapolated.
\label{fig_3S1}}
\end{figure}

\begin{figure}[t]
\centering
\includegraphics[width = \columnwidth]{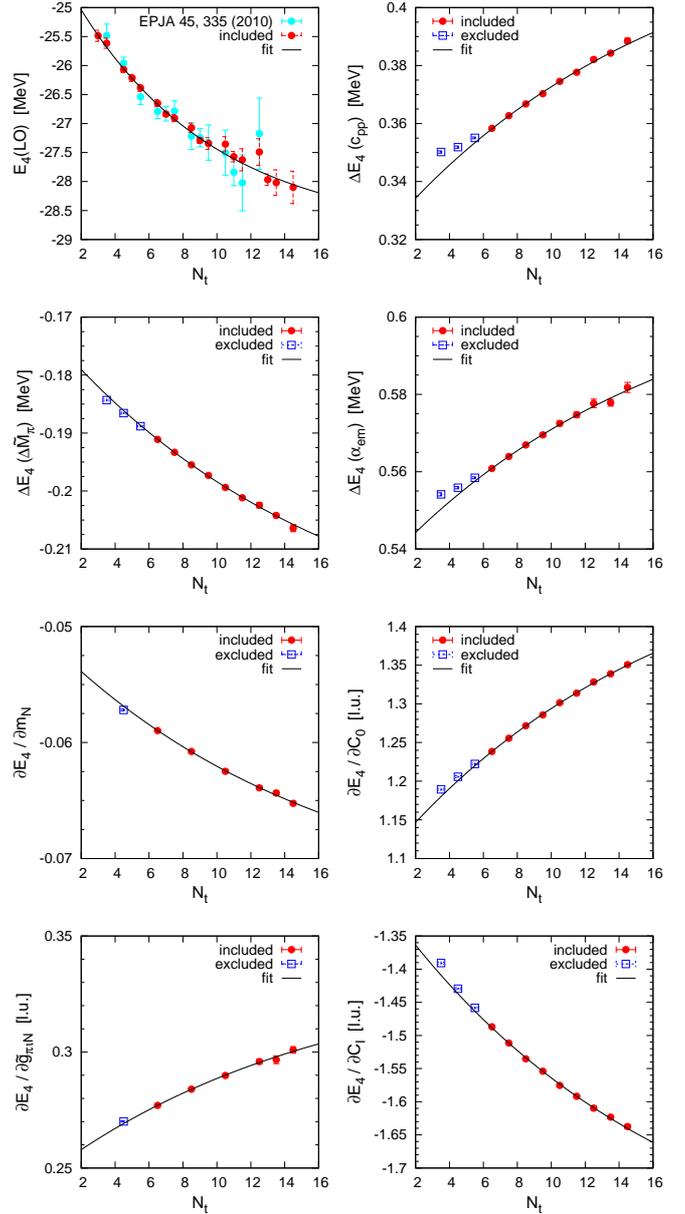}
\caption{AFQMC calculation of $^4$He, as a function of Euclidean time steps $N_t^{}$.
Results after extrapolation $N_t^{} \to \infty$ are given in Table~\ref{tab_48}.
The results of Ref.~\cite{Epelbaum:2010xt} for $E_4^{}$(LO) are included 
to highlight the improved statistics, and as a consistency check.
\label{fig_4He}}
\vspace{-3mm}
\end{figure}

\begin{figure}[t]
\centering
\includegraphics[width = \columnwidth]{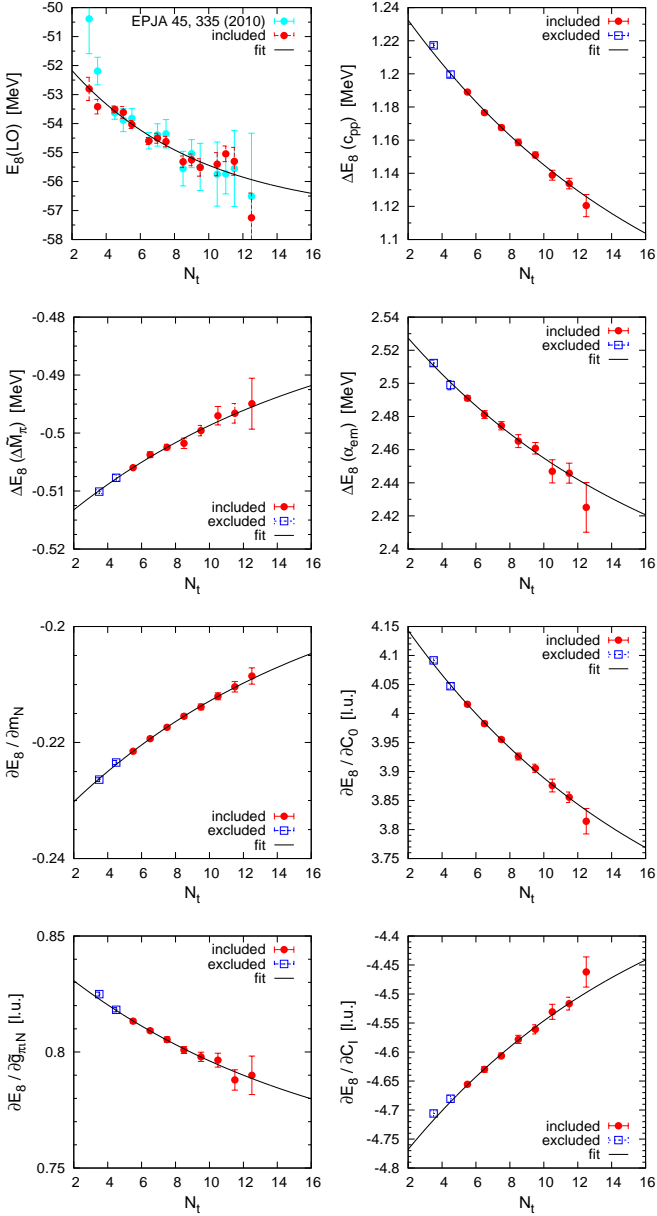}
\caption{AFQMC calculation of $^8$Be, as a function of Euclidean time steps $N_t^{}$.
Results after extrapolation $N_t^{} \to \infty$ are given in Table~\ref{tab_48}.
The results of Ref.~\cite{Epelbaum:2010xt} for $E_8^{}$(LO) are included 
to highlight the improved statistics, and as a consistency check.
\label{fig_8Be}}
\vspace{-3mm}
\end{figure}

\begin{figure}[t]
\centering
\includegraphics[width = \columnwidth]{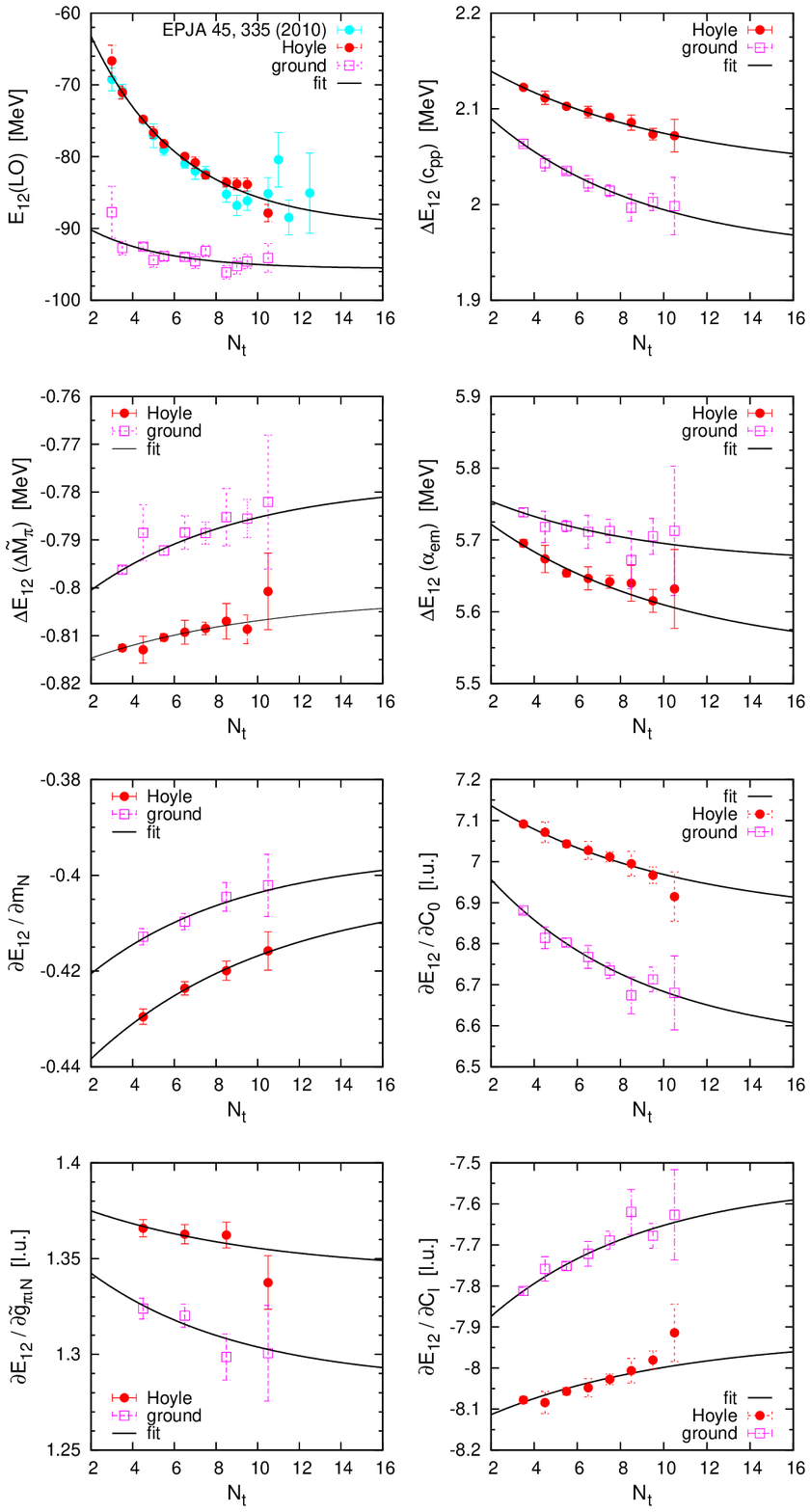}
\caption{AFQMC calculation of $^{12}$C, as a function of Euclidean time steps $N_t^{}$.
Results after extrapolation $N_t^{} \to \infty$ are given in Table~\ref{tab_12C}.
The results of Ref.~\cite{Epelbaum:2010xt} for $E_{12}^\star$(LO) are included 
to highlight the improved statistics, and as a consistency check.
\label{fig_12C}}
\vspace{-3mm}
\end{figure}

We shall first provide an argument for the reliability of our single-exponential description by considering the 
extrapolation to $N_t^{} = \infty$ for the deuteron. Such a validation of our extrapolation procedure is given in Table~\ref{tab_deut}, 
where the extrapolated AFQMC results for the deuteron
(see Fig.~\ref{fig_3S1}) are compared with a direct numerical solution of the 
Schr\"odinger equation for an identical Hamiltonian.
The AFQMC test results for the deuteron were obtained in a relatively small
box size of $L = 5.92~\mathrm{fm}$ ($N = 3$). The comparison
in Table~\ref{tab_deut} shows excellent agreement between the extrapolated and 
exact results, which gives confidence that the 
extrapolation procedure employed in our analysis is indeed reliable. We may 
then proceed with the AFQMC calculation for the nuclei relevant
to the triple-alpha process. The results for $^4$He and $^8$Be are summarized
in Table~\ref{tab_48}, and the individual MC data points along with the
extrapolation are shown in Figs.~\ref{fig_4He} and~\ref{fig_8Be}. The results 
for the ground and Hoyle states of $^{12}$C are similarly given in
Table~\ref{tab_12C} and Fig.~\ref{fig_12C}. These AFQMC calculations 
were performed on a $L = 11.84~\mathrm{fm}$ ($N = 6$) lattice, which is
large enough to render residual finite volume effects smaller than the 
expected error from truncation of the chiral expansion of the NN interaction 
at N$^2$LO.

\begin{table}[t]
\centering{
\caption{AFQMC results for $^4$He and $^8$Be. The LO energies $E_i^{}$ and the 
corresponding energy shifts (including the EM shifts) and derivatives (at the 
physical point) have been extrapolated
$N_t^{} \to \infty$. The appropriate units are given for each quantity, 
with ``[l.u.]'' indicating units of the inverse (spatial) lattice spacing. 
Parentheses indicate one-standard-deviation errors. All derivatives are
computed at the physical point.
\label{tab_48}}
\vspace{.3cm}
\begin{tabular}{|l|r|r|}
\hline
& \multicolumn{1}{c|}{ $^4$He (MC+ex)}
& \multicolumn{1}{c|}{ $^8$Be (MC+ex)}
\\ \hline
$E_i^{} (\mathrm{LO})$ [MeV]
& $-28.89(11)$ & $-57.2(5)$ \\
$\Delta E_i^{} (\Delta \tilde M_\pi^{})$ [MeV]
& $-0.2290(17)$ & $-0.477(5)$ \\  
$\Delta E_i^{} (c_{pp}^{})$ [MeV]
& $0.433(3)$ & $1.02(3)$ \\
$\Delta E_i^{} (\alpha_\mathrm{em}^{})$ [MeV]
& $0.613(2)$ & $2.35(2)$ \\ 
$\partial E_i^{} / \partial m_N^{}$
& $-0.0750(7)$ & $-0.187(6)$ \\
$\partial E_i^{} / \partial \tilde g_{\pi N}^{}$ [l.u.]
& $0.337(3)$ & $0.746(12)$ \\
$\partial E_i^{} / \partial C_0^{}$ [l.u.]
& $1.527(12)$ & $3.52(8)$ \\
$\partial E_i^{} / \partial C_I^{}$ [l.u.] 
& $-1.881(17)$ & $-4.22(7)$\\
\hline
\end{tabular}
}
\end{table}

\begin{table}[t]
\centering{
\caption{AFQMC results for the Hoyle state (second column) and the
  ground state (third column) of $^{12}$C. The LO energies $E_i^{}$ and 
the corresponding energy shifts (including the EM shifts) and derivatives 
(at the physical point) have been extrapolated
$N_t^{} \to \infty$. The appropriate units are given for each quantity, 
with ``[l.u.]'' indicating units of the inverse (spatial) lattice spacing. 
Parentheses indicate one-standard-deviation errors. 
All derivatives are computed at the physical point.
\label{tab_12C}}
\vspace{.3cm}
\begin{tabular}{|l|r|r|}
\hline
& \multicolumn{1}{c|}{ $^{12}$C$^\star$(MC+ex)}
& \multicolumn{1}{c|}{ $^{12}$C (MC+ex)}
\\ \hline
$E_i^{} (\mathrm{LO})$ [MeV]
& $-89.8(13)$ & $-95.6(6)$ \\
$\Delta E_i^{} (\Delta \tilde M_\pi^{})$ [MeV]
& $-0.802(2)$ & $-0.778(4)$ \\  
$\Delta E_i^{} (c_{pp}^{})$ [MeV]
& $2.032(10)$ & $1.95(2)$ \\
$\Delta E_i^{} (\alpha_\mathrm{em}^{})$ [MeV]
& $5.54(2)$ & $5.67(2)$ \\ 
$\partial E_i^{} / \partial m_N^{}$
& $-0.403(5)$ & $-0.395(5)$ \\
$\partial E_i^{} / \partial \tilde g_{\pi N}^{}$ [l.u.]
& $1.343(13)$ & $1.285(16)$ \\
$\partial E_i^{} / \partial C_0^{}$ [l.u.]
& $6.86(3)$ & $6.55(7)$ \\
$\partial E_i^{} / \partial C_I^{}$ [l.u.] 
& $-7.92(3)$ & $-7.54(6)$\\
\hline 
\end{tabular}
}
\end{table}


We may now combine our AFQMC results with the two-nucleon scattering analysis, 
in order to obtain predictions for the $M_\pi^{}$-dependence of the various
states featuring in the triple-alpha process. This can be
performed straightforwardly by substituting the AFQMC results and the numbers 
from Table~\ref{tab_scatt_24} into Eq.~(\ref{QR}),
and by propagating the various sources of error. As described in
Section~\ref{sec:NN}, we adopt
\begin{equation}
x_1^{} = 0.73_{-0.16}^{+0.24}\,, \quad \quad
x_2^{} = -0.024_{-0.034}^{+0.032} \mbox{ l.u.,} 
\end{equation}
for the central values and uncertainties in
$x_1^{}$ and $x_2^{}$, and express the results as a function of $\bar A_s^{}$ 
and $\bar A_t^{}$. In this way, we obtain the following results for the 
$M_\pi^{}$-dependence of the energy levels involved in the triple-alpha process,
\begin{align}
\left. 
\frac{\partial E_4^{}}{\partial M_\pi^{}} 
\right|_{M_\pi^\mathrm{ph}} = &
- 0.339(5) \, \bar A_s^{}
- 0.697(4) \, \bar A_t^{} 
\nonumber \\
& + 0.0380(14) {^{+0.008}_{-0.006}},
\label{resultE_4} \\
\left. 
\frac{\partial E_8^{}}{\partial M_\pi^{}} 
\right|_{M_\pi^\mathrm{ph}} = &
- 0.794(32) \, \bar A_s^{}
- 1.584(23) \, \bar A_t^{}
\nonumber \\
& + 0.089(9){^{+0.017}_{-0.011}},
\label{resultE_8} \\
\left. 
\frac{\partial E_{12}^{}}{\partial M_\pi^{}} 
\right|_{M_\pi^\mathrm{ph}} = &
- 1.52(3) \, \bar A_s^{}
- 2.88(2) \, \bar A_t^{}
\nonumber \\
& + 0.159(7){^{+0.023}_{-0.018}}, 
\label{resultE_12} \\
\left. 
\frac{\partial E_{12}^\star}{\partial M_\pi^{}} 
\right|_{M_\pi^\mathrm{ph}} = &
- 1.588(11) \, \bar A_s^{}
- 3.025(8) \, \bar A_t^{} 
\nonumber \\
&+ 0.178(4){^{+0.026}_{-0.021}}, 
\label{resultE_12s}
\end{align}
where the error receives contributions both from the statistical
error of the AFQMC calculation (given in parentheses) as well as from 
the uncertainties in $x_1^{}$ and $x_2^{}$ (explicit positive and negative 
bounds given). It is noteworthy that $x_1^{}$ and $x_2^{}$ only affect the 
constant terms in the above results, and therefore the (sizable) uncertainties 
in these coefficients have a relatively minor impact.
We can also assess the sensitivity to small shifts in $M_\pi^{}$ by computing 
the ``$K$-factors'' as defined in Eq.~(\ref{Kfact}).
For this purpose, we take $M_\pi^{} = 138.0$~MeV as the isospin-averaged pion
mass, and the empirical values $E_4^{\mathrm{exp}} = -28.30$~MeV,
$E_8^{\mathrm{exp}} = -56.50$~MeV,  $E_{12}^{\mathrm{exp}} = -92.16$~MeV, and 
$E_{12}^{\star\mathrm{exp}} = -84.51$~MeV for the $E_i^{}$. 
This yields
\begin{align}
K^\pi_{E_4^{}} = & \:\: 
1.652(25) \, \bar A_s^{}
+ 3.401(21) \, \bar A_t^{}
- 0.185(7){^{+0.029}_{-0.039}}, \label{resultK_e4}
\\
K^\pi_{E_8^{}} = & \:\: 
1.94(8) \, \bar A_s^{}
+ 3.87(6) \, \bar A_t^{}
- 0.217(21){^{+0.027}_{-0.041}}, \label{resultK_e8}
\\
K^\pi_{E_{12}^{}} = & \:\: 
2.27(4) \, \bar A_s^{}
+ 4.32(3) \, \bar A_t^{}
- 0.239(11){^{+0.026}_{-0.034}}, \label{resultK_e12}
\\
K^\pi_{E_{12}^\star} = & \:\: 
2.593(19) \, \bar A_s^{} 
+ 4.940(13) \, \bar A_t^{} 
- 0.291(7){^{+0.034}_{-0.043}}, \label{resultK_e12s}
\end{align}
where the same conventions for the errors have been applied. Having 
calculated the shifts of the individual energy levels involved in the 
triple-alpha process, we may combine these and obtain similar predictions for 
the energy differences $\Delta E_b^{}$, $\Delta E_h^{}$ and in particular for 
$\varepsilon$,  which is the critical control parameter for the triple-alpha 
reaction rate $r_{3\alpha}^{}$ in Eq.~(\ref{rate}). We find
\begin{align}
\left. 
\frac{\partial \Delta E_b^{}}{\partial M_\pi^{}} 
\right|_{M_\pi^\mathrm{ph}} = &  
- 0.117(34) \, \bar A_s^{}
- 0.189(24) \, \bar A_t^{}
\nonumber \\
& + 0.013(9){^{+0.003}_{-0.002}}, \label{result_eb}
\\
\left. 
\frac{\partial \Delta E_h^{}}{\partial M_\pi^{}} 
\right|_{M_\pi^\mathrm{ph}} = &
- 0.455(35) \, \bar A_s^{}
- 0.744(24) \, \bar A_t^{}
\nonumber \\
& + 0.051(10){^{+0.008}_{-0.009}}, \label{result_eh}
\\
\left. 
\frac{\partial \varepsilon}{\partial M_\pi^{}} 
\right|_{M_\pi^\mathrm{ph}} = &
- 0.572(19) \, \bar A_s^{}
- 0.933(15) \, \bar A_t^{}
\nonumber \\
& + 0.064(6){^{+0.010}_{-0.009}}, \label{result_eps}
\end{align}
and
\begin{align}
K^\pi_{\Delta E_b^{}} &= 
- 175(51) \, \bar A_s^{}
- 284(36) \, \bar A_t^{}
+ 19(13){^{+4.5}_{-3.0}}, \label{resultK_eb}
\\
K^\pi_{\Delta E_h^{}} &= 
- 217(16) \, \bar A_s^{}
- 355(12) \, \bar A_t^{}
+ 25(5){^{+4.0}_{-4.5}}, \label{resultK_eh} 
\\
K^\pi_{\varepsilon} &= 
- 208(7) \, \bar A_s^{}
- 339(5) \, \bar A_t^{}
+ 23(2){^{+3.7}_{-3.4}}, \label{resultK_e}
\end{align}
where we have used the empirical values $\Delta E_{b}^{\rm exp}=92$~keV, 
$\Delta E_{h}^{\rm exp}=289$~keV, and $\varepsilon = 380$~keV. 


We now turn our attention to the AFQMC results for the EM shifts. As explained 
in Section~\ref{sec:anatomy}, our objective is to compute the energy shifts 
$Q_\mathrm{em}^{}(E_i^{})$ defined in Eq.~(\ref{defQ}). This involves the
unknown parameter $x_{pp}^{}$, which determines the relative strength of 
the proton-proton contact interaction that emerges from the lattice
regularization of the long-range Coulomb force.
We may fix $x_{pp}^{}$ by means of the known contribution of the Coulomb force 
to the binding energy of $^4$He, which is
$Q_\mathrm{em}^{}(E_4^{}) = 0.78(3)$ MeV~\cite{Nogga}. The quoted error
reflects the model-dependence and is determined from the range of values 
corresponding to different phenomenological two- and
three-nucleon potentials as well as nuclear forces derived in chiral EFT. 
Using the AFQMC results for $\Delta E_i^{}(\alpha_\mathrm{em}^{})$ and 
$\Delta E_i^{}(c_{pp}^{})$ from Table~\ref{tab_48}, 
we find
\begin{align}
Q_\mathrm{em}^{}(E_4^{}) & = 0.433(3)~\mathrm{MeV} \times x_{pp}^{} + 0.613(2)~\mathrm{MeV} 
\nonumber \\
& \stackrel{!}{=} {0.78(3)~\mathrm{MeV}} \:\: \to \:\:
x_{pp}^{} \simeq 0.39(5),
\end{align}
for $^4$He, which enables us to predict
\begin{align}
Q_\mathrm{em}^{}(E_8^{}) & = 1.02(3)~\mathrm{MeV} \times x_{pp}^{} + 2.35(2)~\mathrm{MeV} 
\nonumber \\
& = 2.75(8)~\mathrm{MeV},
\end{align}
for $^8$Be. Further, using the value $x_{pp}^{} = 0.39(5)$ and the AFQMC results given in Table~\ref{tab_12C}, we
predict
\begin{align}
Q_\mathrm{em}^{}(E_{12}^\star) & = 2.032(10)~\mathrm{MeV} \times x_{pp}^{} + 5.54(2)~\mathrm{MeV} 
\nonumber \\
& = 6.33(6)~\mathrm{MeV},
\end{align}
for the Hoyle state, and
\begin{align}
Q_\mathrm{em}^{}(E_{12}^{}) & = 1.95(2)~\mathrm{MeV} \times x_{pp}^{} + 5.67(2)~\mathrm{MeV} 
\nonumber \\
&= 6.43(6)~\mathrm{MeV},
\end{align}
for the ground state of $^{12}$C. We are now in the position to 
predict the EM shifts of the energy differences $\Delta E_b^{}$, 
$\Delta E_h^{}$ and $\varepsilon$. This gives
\begin{equation}
\begin{split}
Q_\mathrm{em}^{}(\Delta E_h^{}) &  = 2.80(10)~\mathrm{MeV}, \\
Q_\mathrm{em}^{}(\Delta E_b^{}) & = 1.19(8)~\mathrm{MeV}, \\
Q_\mathrm{em}^{}(\varepsilon) & = 3.99(9)~\mathrm{MeV}.
\label{Qfacts}
\end{split}
\end{equation}


We have also studied the dependence of the excitation energy 
$\Delta E_c^{}$ of the Hoyle state on $M_\pi^{}$ and $\alpha_\mathrm{em}^{}$. 
Although $\Delta E_c^{}$ is not needed for the calculation of the triple-alpha 
reaction rate, it provides an instructive
reference point when discussing the sensitivity of various energy differences 
to small changes in the fundamental constants. We compute $\Delta E_c^{}$ from 
\begin{equation}
\Delta E_c^{} \equiv  E_{12}^\star - E_{12}^{},
\end{equation}
for which we find
\begin{align}
\left. \frac{\partial \Delta E_c^{}}{\partial M_\pi^{}}
\right|_{M_\pi^\mathrm{ph}} = &
- 0.07(3) \, \bar A_s^{}
- 0.14(2) \, \bar A_t^{}
\nonumber \\
& + 0.019(9){^{+0.004}_{-0.003}}~.
\end{align}
This yields
\begin{equation}
K^\pi_{\Delta E_c^{}} =
- 1.3(5) \, \bar A_s^{}
- 2.6(4) \, \bar A_t^{}
+ 0.34(15){^{+0.07}_{-0.05}}~,
\label{resultK_ec}
\end{equation}
for the sensitivity to small changes in $M_\pi^{}$. The above result 
corresponds to the  empirical  value of 
$\Delta E_c^{\mathrm{exp}} = 7.65$~MeV. Finally, we find
\begin{equation}
Q_\mathrm{em}^{}(\Delta E_c^{}) = 0.10(7)~\mathrm{MeV},
\end{equation}
for the EM shift in the excitation energy of the Hoyle state.


\section{Theoretical uncertainties and higher-order corrections} 
\label{sec:HigherOrders}

In order to estimate the theoretical uncertainty of our results, we shall 
first consider the effects of neglected higher-order terms in the chiral 
EFT expansion. To this end, we compute the
$M_\pi^{}$-dependence of the $^4$He binding energy induced by the 
explicitly $M_\pi^{}$-dependent part of the 
three-nucleon force~(3NF), 
\begin{align}
\mathcal{A}_{\rm N^2LO}^{\rm 3NF} = & \:\: \frac{g_A^2}{8 F_\pi^4} \, 
\frac{\vec q_i^{} \cdot \vec \sigma_i^{} \: \vec q_j^{} \cdot \vec \sigma_j^{}}
{(\vec q_i^2 + M_\pi^2) (\vec q_j^2 + M_\pi^2)} 
\Big[ \, \fet \tau_i^{} \cdot \fet \tau_j^{} \, (-4 c_1^{} M_\pi^2
\nonumber \\
& + 2 c_3^{} \, \vec q_i^{} \cdot \vec q_j^{}) + c_4^{} (\fet \tau_i^{} \times \fet \tau_j^{}) \cdot
\fet \tau_k^{} \: (\vec q_i^{} \times \vec q_j^{} ) \cdot \vec q_k^{} \Big]
\nonumber \\
& - \frac{g_A^{} D}{8 F_\pi^2} \,
\frac{\vec q_i^{} \cdot \vec \sigma_i^{} \: \vec q_i^{} \cdot \vec \sigma_j^{} 
\: \fet \tau_i^{} \cdot \fet \tau_j^{}
}{\vec q_i^2 + M_\pi^2} 
+ \frac{E}{2} \, \fet \tau_i^{} \cdot \fet \tau_j^{}  
\nonumber \\
& + \mbox{permutations},
\label{3NF}
\end{align}
which contributes at N$^2$LO. 
For the LECs $c_i^{}$, we take $c_1^{} 
= -0.81$~GeV$^{-1}$, $c_3^{} = -4.7$~GeV$^{-1}$, and $c_4^{} 
= 3.4$~GeV$^{-1}$, as determined from low-energy pion-nucleon scattering~\cite{Bernard:1995dp,Buettiker:1999ap}. 
The LECs $D$ and $E$ are fixed by means of the triton binding energy and 
the weak axial vector current. For more details on the treatment of the 3NF in the nuclear
lattice simulations, see Ref.~\cite{Epelbaum:2009zsa}.  

It should be understood that the $M_\pi^{}$-dependence of the $E_i^{}$ induced by that of the 
3NF is beyond the accuracy of our analysis. The following estimate of its impact on 
$E_4^{}$ is only intended as a consistency check.
We calculate the sensitivity of $E_4^{}$ to changes in $M_\pi^{}$ entering the 3NF by performing AFQMC
calculations using a slightly shifted pion mass in Eq.~(\ref{3NF}),  
namely $\tilde M_\pi^{} = M_\pi^{\rm ph} \pm 20$~MeV. The resulting shifts in
$E_4^{}$ induced by the 3NF are 
\begin{align}  
\Delta E_4^\mathrm{3NF} \big|_{\tilde M_\pi^{} 
= M_\pi^{\rm ph} + \, 20~\mathrm{MeV}} = -3.034(12)~\mathrm{MeV}, \nonumber \\
\Delta E_4^\mathrm{3NF} \big|_{\tilde M_\pi^{} 
= M_\pi^{\rm ph} - \, 20~\mathrm{MeV}} = -3.204(16)~\mathrm{MeV},
\end{align}
which gives us the rough estimate
\begin{equation}
\left. \frac{\partial E_4^\mathrm{3NF}}{\partial \tilde M_\pi^{}} 
\right |_{M_\pi^\mathrm{ph}} \simeq 0.004~.
\end{equation}
This is an order of magnitude smaller than e.g.~the LO contribution
$\partial E_4^{}/\partial \tilde M_\pi^{} |_{M_\pi^\mathrm{ph}} \simeq -0.05$.
This observation suggests that the 3NF effects are indeed very much 
suppressed, as expected for a N$^2$LO contribution. 


It is also instructive to compare our results with the ones of
Ref.~\cite{Bedaque:2010hr}, which were obtained within the framework 
of \emph{pionless} EFT. Not only does this provide a useful consistency 
check for our calculations, but it also 
allows us to estimate our theoretical uncertainty in a complementary way. 
In particular, we can compare our result for $\partial E_4^{} / 
\partial M_\pi^{}$ in Eq.~(\ref{resultE_4}) with the result for the $^4$He
binding energy $B_4^{}$ ($B_i^{} \equiv | B_i^{} | = - E_i^{}$) 
given in Eqs.~(1.5)-(1.7) of Ref.~\cite{Bedaque:2010hr}
\begin{equation}
\label{BLPorig}
\frac{\partial B_4^{}}{\partial m_q^{}} \simeq 0.037 \, \frac{B_4^{}}{a_s^{}}
\frac{\partial a_s^{}}{\partial m_q^{}} + 0.74 \, \frac{B_4^{}}{B_d^{}}
\frac{\partial B_d^{}}{\partial m_q^{}},
\end{equation} 
where $B_d^{}$ is the deuteron binding energy. 
Noting that 
$\partial/\partial m_q^{} \propto \partial/\partial M_\pi^{}$ and making use of the relation
\begin{equation}
\frac{\partial a_s^{}}{\partial M_\pi^{}} = -a_s^2 \, \frac{\partial a_s^{-1}}{\partial M_\pi^{}},
\end{equation}
we can bring Eq.~(\ref{BLPorig}) into the form 
\begin{equation}
\label{BLPinterm}
\frac{\partial B_4^{}}{\partial M_\pi^{}} \simeq -0.037 \, B_4^{} a_s^{} 
\frac{\partial a_s^{-1}}{\partial m_q^{}} + 0.74 \, \frac{B_4^{}}{B_d^{}}
\frac{\partial B_d^{}}{\partial M_\pi^{}},
\end{equation} 
where it is still necessary to convert the dependence on $B_d^{}$ into 
a corresponding dependence on $a_t^{}$.
To this end, we use the effective range approximation 
\begin{equation}
p \cot \delta_t^{} \simeq - \frac{1}{a_t^{}} + \frac{1}{2} \, p^2 r_t^{},
\end{equation}
to obtain 
\begin{align} 
\frac{\partial B_d^{}}{\partial M_\pi^{}} & \simeq \frac{4}{m_N^{} [a_t^{} +
\sqrt{a_t^{}(a_t^{} - 2 r_t^{})} - 2 r_t^{}]} \, \frac{\partial a_t^{-1}}{\partial M_\pi^{}}  \nonumber \\
& \simeq 0.164 \, \frac{\partial a_t^{-1}}{\partial M_\pi^{}},
\end{align}
where we have inserted the empirical values $a_t^{} = 5.42$~fm and $r_t^{} = 1.75$~fm.
Finally, by combining this expression with Eq.~(\ref{BLPinterm}) and noting 
that $B_i^{} = -E_i^{}$, the  pionless EFT result of
Ref.~\cite{Bedaque:2010hr} is brought into the form  
\begin{align}
\left. \frac{\partial E_4^{}}{\partial M_\pi^{}} \right|_{M_\pi^\mathrm{ph}}
\simeq & -0.037 \, E_4^{} a_s^{} \bar A_s^{}
- 1.48 \, \frac{E_4^{}}{E_d^{} m_N^{} a_t^{}} \, \bar A_t^{}  \nonumber \\
\simeq & -0.126 \, \bar A_s^{} - 0.739 \, \bar A_t^{},
\label{BLPfinal}
\end{align}
which is in a reasonable agreement with Eq.~(\ref{resultE_4}). 
By taking the central values of the coefficients $x_1^{}$ and $x_2^{}$ in 
Eq.~(\ref{resultE_4}), as well as $\bar A_s^{} = 0.29$ and $\bar A_t^{} =
-0.18$ as found in the most recent chiral EFT calculations (see Section~\ref{sec:discussion} 
for details), we obtain $\partial E_4^{} / \partial M_\pi^{}  \simeq 0.065$. 
This should be compared with the pionless EFT result  $\partial E_4^{} /
\partial M_\pi^{} \simeq 0.096$ based on Eq.~(\ref{BLPfinal}). 
We expect that the theoretical uncertainty of our calculation is much smaller 
than the difference between these two numbers or, more generally, than 
the difference  between Eqs.~(\ref{BLPfinal}) and~(\ref{resultE_4}).
This is because in our approach, the uncertainty is entirely due to
suppressed higher order corrections.

Finally, we note that our analysis does not account for the 
$M_\pi^{}$-dependence of the momentum-dependent, sub-leading contact 
interactions in the chiral EFT Hamiltonian. Assuming validity of the naive 
dimensional analysis, such effects are beyond the accuracy 
of our present work. In addition, the strong correlations we observe for the 
$M_\pi^{}$-dependence of the various observables (as discussed in
Section~\ref{sec:correlations}) indicate that the relevant 
dynamics is largely governed by the large $S$--wave NN scattering lengths. 
Higher-order $M_\pi^{}$-dependent  short-range terms are therefore expected 
to play a lesser role. 


\section{Correlations and the binding energy of the alpha particle}
\label{sec:correlations}

Given the results in Section~\ref{sec:AFQMC}, we are now in a position to 
draw conclusions concerning the individual energies $E_i^{}$ and the
associated energy differences. The first interesting observation
is that the energy differences $\Delta E_h^{}$, $\Delta E_b^{}$ and 
$\varepsilon$ are, by themselves, extremely sensitive to changes in 
$M_\pi^{}$ as could be expected. Such a conclusion follows from the
unnaturally large coefficients in Eqs.~(\ref{resultK_eb})-(\ref{resultK_e}).
Notice that the fact that $\Delta E_h^{}$, $\Delta E_b^{}$ and $\varepsilon$
are much smaller than the individual $E_i^{}$ does not, by itself, imply a 
strong fine-tuning. For example, the sensitivity of the Hoyle state excitation 
energy $\Delta E_c^{}$ in Eq.~(\ref{resultK_ec}) is
of a natural size, in spite of $\Delta E_c^{\rm exp} = 7.65$~MeV being almost an
order of magnitude smaller than $| E_{12}^{\rm exp} | = 92.16$~MeV.   

While we find that $\Delta E_h^{}$, $\Delta E_b^{}$ and $\varepsilon$ are by 
themselves extremely sensitive to variations in $M_\pi^{}$, we also observe 
the approximate relations
\begin{equation}
\frac{K_{\Delta E_h^{}}^{\pi}}{K_{\Delta E_b^{}}^{\pi}} \simeq 1.25,
\quad\quad
\frac{K_{\Delta E_h^{}}^{\pi}}{K_{\varepsilon}^{\pi}} \simeq 1.05,
\end{equation}
which are satisfied for the central values of the individual terms in 
Eqs.~(\ref{resultK_eb})-(\ref{resultK_e}) at the level of a few percent. 
This suggests that $\Delta E_h^{}$, $\Delta E_b^{}$ and $\varepsilon$
cannot be independently varied (or fine-tuned) by changing the singlet 
and triplet NN scattering lengths 
(or, equivalently, by changing the strength of the short-range NN force 
in the $^1S_0$ and $^3S_1$ channels). 
Moreover, it is apparent from Eqs.~(\ref{resultK_e4})-(\ref{resultK_e12s}) that 
also  the energies $E_i^{}$ of the individual states are 
as well strongly correlated in a similar manner.

\begin{figure}[t]
\centering
\includegraphics[trim = 0cm 0cm 0cm 0cm, clip, width = .9\columnwidth]{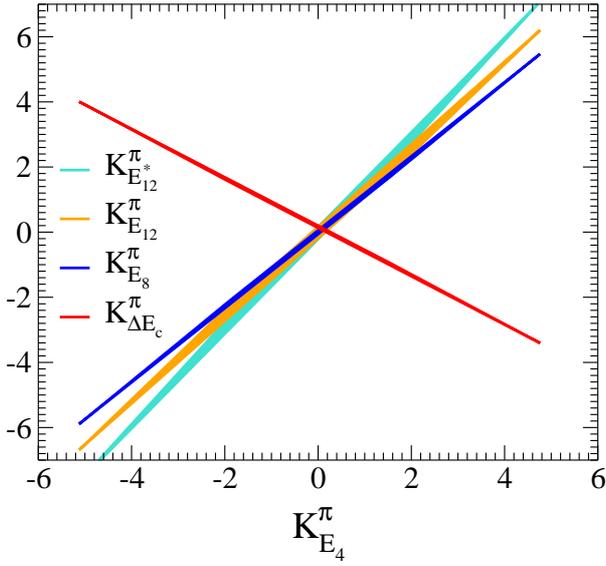}
\caption{Sensitivities of $E_8^{}$, $E_{12}^{}$, $E_{12}^\star$ and $\Delta
  E_c^{}$ to changes in $M_\pi^{}$, as a function of $K_{E_4^{}}^\pi$ under 
independent variation of $\bar A_s^{}$ and $\bar A_t^{}$ over the range 
$\{-1 \ldots 1\}$.    The bands correspond to $E_{12}^\star$,  $E_{12}^{}$,  
$E_8^{}$ and $\Delta E_c^{}$ in clockwise order.
\label{fig_corr1}}
\end{figure}

\begin{figure}[t]
\centering
\includegraphics[trim = 0cm 0cm 0cm 0cm, clip, width = .9\columnwidth]{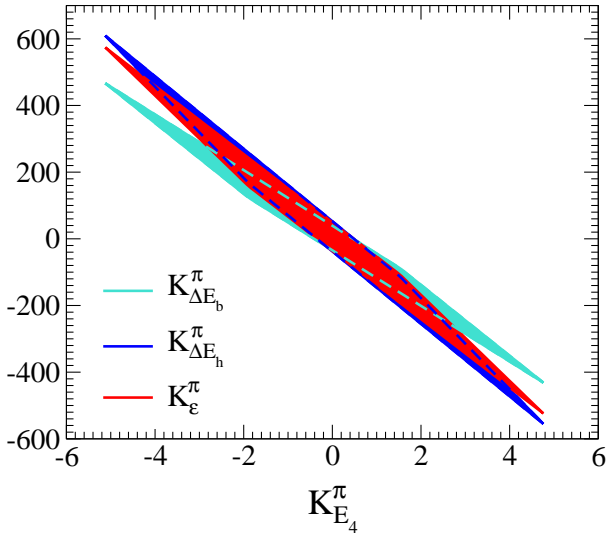}
\caption{Sensitivities of $\Delta E_h^{}$, $\Delta E_b^{}$ and $\varepsilon$ 
to changes in $M_\pi^{}$, as a function of $K_{E_4^{}}^\pi$ under independent 
variation of $\bar A_s^{}$ and $\bar A_t^{}$ over the range $\{-1 \ldots 1\}$.   
The bands correspond to $\Delta E_b^{}$, 
$\varepsilon$ and  $\Delta E_h^{}$ in clockwise order.
\label{fig_corr2}}
\vspace{-3mm}
\end{figure}

In order to quantify and illustrate the observed correlations, we show in 
Fig.~\ref{fig_corr1} the changes in the sensitivities of $E_8^{}$,
$E_{12}^{}$, $E_{12}^\star$ and $\Delta E_c^{}$ as a function of 
$K_{E_4^{}}^\pi$, when $\bar A_s^{}$ and $\bar A_t^{}$ are independently
varied over a large range. The correlations for
$\Delta E_h^{}$, $\Delta E_b^{}$ and $\varepsilon$ are shown in a similar way 
in Fig.~\ref{fig_corr2}. Within the statistical accuracy of our AFQMC results, 
we may conclude that the scenario of independent variations of the energy 
levels pertinent to the triple-alpha process under changes in the
fundamental parameters is strongly disfavored. 
Given the prominent role of the $^4$He binding energy in the correlations, the observed
behavior is strongly suggestive of the $\alpha$-cluster structure of
the $^8$Be, $^{12}$C and Hoyle states.  Such
correlations related to the production of carbon have 
indeed been speculated upon earlier~\cite{Livio,WeinbergFacing}.  

\section{Reaction rate of the triple-alpha process}
\label{sec:discussion}

\begin{figure}[t]
\includegraphics[trim = 0cm 0cm 0cm 0cm, clip, width = \columnwidth]{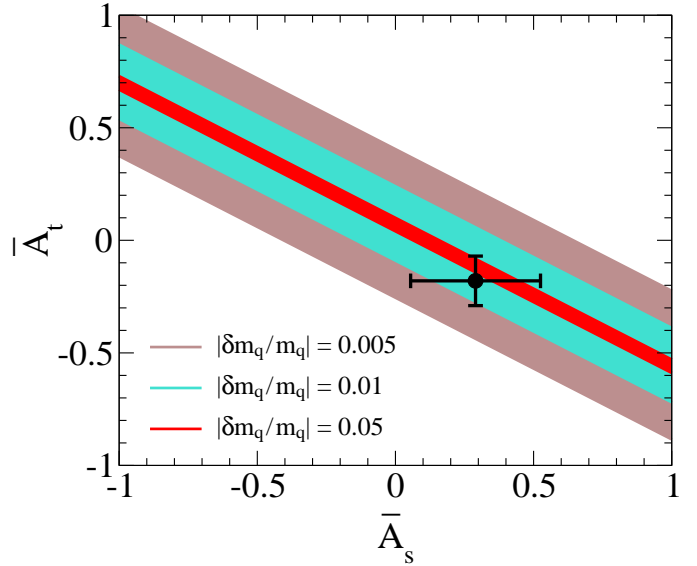}
\caption{``Survivability bands'' for carbon-oxygen based life from
  Eq.~(\ref{final_res}), due to  $0.5\%$ (broad outer band), $1\%$ (medium
  band) and $5\%$ (narrow inner band) changes in $m_q^{}$ in terms of the
  input parameters $\bar A_s^{}$ and $\bar A_t^{}$. The most up-to-date 
  N$^2$LO analysis of $\bar A_s^{}$ and $\bar A_t^{}$ corresponding to 
  Eq.~(\ref{Evgeny1}) is given by the data point with horizontal and vertical error bars.     
\label{fig_end}}
\vspace{-3mm}
\end{figure}

We now turn our attention to the reaction rate of the triple-alpha 
process as given by Eq.~(\ref{rate}), and determine the range of 
variations in $m_q^{}$ and $\alpha_{\rm em}^{}$ compatible with the formation 
of significant amounts of carbon and oxygen
in our Universe, and thus with the existence of carbon-oxygen based life. We
recall that the stellar modeling calculations of 
Refs.~\cite{Oberhummer:2000zj,Oberhummer:1999ab} suggest that sufficient 
abundances of both carbon and oxygen can be maintained within an envelope of 
$\pm 100$~keV around the empirical value of $\varepsilon = 379.47(18)$~keV.
For small variations $| \delta \alpha_{\rm em}^{}/\alpha_{\rm em}^{} | \ll 1$ 
and $| \delta m_q^{}/m_q^{} | \ll 1$, 
the resulting change in $\varepsilon$ can be expressed as
\begin{align} 
\delta (\varepsilon ) \approx &
\left. \frac{\partial \varepsilon}{\partial M_\pi^{}} \right |_{M_\pi^\mathrm{ph}} \delta M_\pi^{} + 
\left. \frac{\partial \varepsilon}{\partial \alpha_{\rm em}^{}} \right |_{\alpha_{\rm em}^\mathrm{ph}} 
\delta \alpha_{\rm em}^{} \\
= & \left. \frac{\partial \varepsilon}{\partial M_\pi^{}} \right|_{M_\pi^\mathrm{ph}}
K_{M_\pi}^q M_\pi^{} \left(\frac{\delta m_q^{}}{m_q^{}}\right) + Q_{\rm em}^{}(\varepsilon) 
\left(\frac{\delta \alpha_{\rm em}^{}}{\alpha_{\rm em}^{}}\right),
\nonumber
\end{align}
where we recall that $K_{M_\pi^{}}^q =
0.494^{+0.009}_{-0.013}$~\cite{Berengut:2013nh}. Thus, 
the  condition $| \delta (\varepsilon )| < 100$~keV together with
Eq.~(\ref{Qfacts}) leads to the predicted tolerance
$| \delta \alpha_{\rm em}^{}/\alpha_{\rm em}^{} | \simeq 2.5 \%$ of 
carbon-oxygen based life to shifts in $\alpha_{\rm em}^{}$.
This result is compatible with the $\simeq 4\%$ bound reported 
in Ref.~\cite{Oberhummer:2000mn}.   For shifts in $m_q^{}$, we find
\begin{align}
& \left| \Big[ 0.572(19) \, \bar A_s^{} + 0.933(15) \, \bar A_t^{} 
- 0.064(6)  \Big]  
\left(\frac{\delta m_q^{}}{m_q^{}} \right) \right| \nonumber \\
& < 0.15\%,
\label{final_res}
\end{align}
using Eq.~(\ref{result_eps}), 
where we have neglected the relatively insignificant errors introduced 
by $x_1^{}$ and $x_2^{}$. The resulting constraints on the values of $\bar
A_s^{}$ and $\bar A_t^{}$ compatible
with the condition $| \delta (\varepsilon )| < 100$~keV are visualized in 
Fig.~\ref{fig_end}.  The various shaded bands in Fig.~\ref{fig_end} cover the 
values of $\bar A_s^{}$ and $\bar A_t^{}$ consistent
with carbon-oxygen based life, when $m_q^{}$ is varied by $0.5$\%, $1$\% and $5$\%.

In the most generic scenario, assuming that both of the dimensionless
quantities $\bar A_s^{}$ and $\bar A_t^{}$ are $\sim {\mathcal O}(1)$,
and therefore $0.572(19) \, \bar A_s^{} + 0.933(15) \, \bar A_t^{} \sim 
\mathcal{O}(1)$, our results imply that a change in $m_q^{}$ of as little
as $\simeq 0.15\%$ would suffice to render carbon-oxygen based life unlikely
to exist. Stated differently, the ``survivability band'' 
corresponding to $|\delta m_q^{} / m_q^{} | < 0.15$\% would cover the whole of 
Fig.~\ref{fig_end}. It should be noted that in such a generic
scenario, one can approximate  
\begin{equation}
\left. \frac{\partial \varepsilon }{\partial M_\pi^{}} \right|_{M_\pi^\mathrm{ph}}
\approx 1.5 \left. \frac{\partial E_4^{}}{\partial M_\pi^{}} \right|_{M_\pi^\mathrm{ph}},
\end{equation}
which implies that the binding energy of $^4$He should be fine-tuned under
variation  of $m_q^{}$ to its empirical 
value at the level of $\simeq 0.25\%$ in order to fulfill the condition 
$| \delta (\varepsilon ) | < 100$~keV.  Nevertheless, there clearly also
exists a special value for the ratio of $\bar A_s^{}$ to $\bar A_t^{}$, given by 
\begin{equation}  
\bar A_s^{}/ \bar A_t^{} \simeq -1.5,
\end{equation}
for which the dependence of $\Delta E_h^{}$, $\Delta E_b^{}$ and $\varepsilon$ 
on $M_\pi^{}$  becomes vanishingly small (compared to the statistical
uncertainties of the AFQMC calculation), such that
the factor
\begin{equation}
0.572(19) \, \bar A_s^{} + 0.933(15) \, \bar A_t^{} - 0.064(6) \ll 1, 
\end{equation}
in Eq.~(\ref{final_res}). In this case, we would conclude that the reaction 
rate of the triple-alpha process were completely
insensitive to shifts in $m_q^{}$. As the realistic scenario is likely to be 
found somewhere in between these extreme cases,
it becomes important to consider the available constraints on $\bar A_s^{}$
and $\bar A_t^{}$ before final conclusions are drawn.

The quark mass dependence of the $S$--wave NN scattering lengths has been
analyzed within the framework of chiral EFT by several groups. 
The problem common to these calculations is the lack of knowledge
about the $m_q^{}$-dependence of the NN contact interactions. Estimating the
size of the corresponding LECs by means of dimensional analysis
typically leads to large uncertainties for chiral extrapolations of
the scattering lengths. For example, the NLO calculation of 
Refs.~\cite{Epelbaum:2002gb,Epelbaum:2002gk} 
resulted in the values
\begin{equation}
K_{a_s^{}}^q =  5 \pm 5, \quad \quad
K_{a_t^{}}^q =  1.1 \pm 0.9,
\label{KfacsOurOld}
\end{equation}
for the relevant $K$--factors. These are consistent with the 
NLO analysis of Ref.~\cite{Beane:2002xf}, which yielded
\begin{equation}
K_{a_s^{}}^q =  2.4 \pm 3.0, \quad \quad
K_{a_t^{}}^q =  3.0 \pm 3.5,
\label{KfacsKSW}
\end{equation}
based on a perturbative treatment of OPE (see Ref.~\cite{Beane:2001bc} 
for a related study). More recently, attempts have been made to combine chiral EFT with 
lattice QCD calculations. In particular, the NPLQCD collaboration
has determined the regions for the $S$--wave scattering lengths
consistent with their lattice results, $a_s^{} = (0.63 \pm 0.50)$~fm
and $a_t^{} = (0.63 \pm 0.74)$~fm, obtained for $M_\pi^{} 
= 353.7 \pm 2.1$~MeV~\cite{Beane:2006mx}. 
By using these lattice data in conjunction with the assumptions of perturbativeness
of the OPE potential in the $^3S_1$--$^3D_1$ channel and validity of the
chiral expansion for NN scattering for  $M_\pi^{} > 350$~MeV, 
Refs.~\cite{Chen:2010yt,Soto:2011tb} obtained results for
$K_{a_s^{}}^q$ and $K_{a_t^{}}^q$ which are consistent with the ones quoted in 
Eq.~(\ref{KfacsOurOld}), and in slight disagreement with those of Eq.~(\ref{KfacsKSW}). 

Very recently, the analysis of the $m_q^{}$-dependence 
of NN observables was extended to N$^2$LO in chiral EFT~\cite{Berengut:2013nh}. 
To overcome the difficulties due to the poorly known $m_q^{}$-dependence
of the short-range NN interactions, the authors of 
Ref.~\cite{Berengut:2013nh} exploited the fact that the LECs 
accompanying the NN contact interactions are saturated by exchanges of 
heavy mesons~\cite{Epelbaum:2001fm}.  By means of a unitarized version of ChPT 
in combination with lattice QCD results, which
describes the $m_q^{}$-dependence of the meson resonances saturating these
LECs, the $m_q^{}$-dependence of the NN observables was analyzed at N$^2$LO
without relying on the chiral expansion of the NN contact interactions. 
The most up-to-date values are then given by
\begin{equation}
\label{Kfinal}
K_{a_s^{}}^q = 2.3^{+1.9}_{-1.8}, \quad \quad 
K_{a_t^{}}^q = 0.32^{+0.17}_{-0.18},
\end{equation}
and
\begin{align}
\label{Evgeny1} 
\bar A_s^{} & = -\frac{1}{a_s^{} M_\pi^{}} \frac{K_{a_s^{}}^q}{K_{M_\pi^{}}^q}   \simeq
0.29^{+0.25}_{-0.23}, \nonumber \\
\bar A_t^{} & = -\frac{1}{a_t^{} M_\pi^{}} \frac{K_{a_t^{}}^q}{K_{M_\pi^{}}^q}   \simeq
 -0.18^{+0.10}_{-0.10},
\end{align}
which are not only consistent with the earlier determinations, but also 
in reasonably good agreement with the (parameter-free) LO
chiral EFT calculation of Ref.~\cite{Epelbaum:2013ij}, which is based on a 
novel, cutoff-independent approach.  Interestingly, direct application of the 
central values in Eq.~(\ref{Evgeny1}) gives $\bar A_s^{} / \bar A_t^{} \simeq -1.6$, 
which leads to a strong cancellation of the dependence of $\Delta E_h^{}$, 
$\Delta E_b^{}$ and $\varepsilon$ on $\bar A_s^{}$ and $\bar A_t^{}$, and
hence also to a mild dependence on $M_\pi^{}$. The range of values 
given in Eq.~(\ref{Evgeny1}) suggests that all contributions to the 
$K$--factors (notably including  $K^\pi_{\Delta E_{c}^{}}$) are of 
$\sim \mathcal{O}(1)$, with the exceptions of $K^\pi_{\Delta E_b^{}}$, $K^\pi_{\Delta E_h^{}}$ and
$K^\pi_{\varepsilon^{}}$.


\section{Summary and conclusions} 
\label{sec:summary}

We may summarize our findings by a brief discussion of Fig.~\ref{fig_end}, 
where we have superimposed the result corresponding
to Eq.~(\ref{Evgeny1}) with the ``survivability bands'' from our 
AFQMC results in Eq.~(\ref{final_res}). Given the current theoretical 
uncertainty in $\bar A_s^{}$ and $\bar A_t^{}$, our results remain compatible 
with a vanishing $\partial \varepsilon / \partial M_\pi^{}$, in other words
with a complete lack of fine-tuning. Interestingly, Fig.~\ref{fig_end} also 
indicates that the triple-alpha process is unlikely to be fine-tuned
to a higher degree than $\simeq 0.8$\% under variation of $m_q^{}$. 
The central values of $\bar A_s^{}$ and $\bar A_t^{}$ from
Eq.~(\ref{Evgeny1}) suggest that variations in the light quark masses 
of up to $2 - 3$\% are unlikely to be catastrophic to the formation of 
life-essential carbon and oxygen. A similar calculation of the tolerance for 
shifts in the EM fine-structure constant $\alpha_{\rm em}^{}$
suggests that carbon-oxygen based life can withstand shifts of 
$\simeq 2.5$\% in $\alpha_{\rm em}^{}$. Beyond such relatively small changes 
in the fundamental parameters, the anthropic principle appears necessary to 
explain the observed abundances of $^{12}$C and $^{16}$O. We also note that the 
fine-tuning in the fundamental parameters is much more severe than the
one in the energy difference $\varepsilon$.

Our \textit{ab initio} lattice calculations account for all sources of quark
mass dependence (explicit as well as implicit) in the LO nuclear 
Hamiltonian in chiral EFT, although we have not performed a strict LO analysis
of the triple-alpha reaction rate. We have considered the potential impact of 
neglected higher-order terms on our results, in particular that of the 3NF 
which starts contributing at N$^2$LO in the chiral expansion, and found that 
our conclusions are likely to be robust against such effects. Therefore, the 
most immediately useful extension of our work would be the incorporation of a more precise
determination of $\bar{A}_s^{}$ and $\bar{A}_t^{}$ from future lattice QCD studies. 

As a longer-term objective, we may envision
the inclusion of dynamical photons in our AFQMC framework. Such a coupling of 
lattice QED to lattice chiral EFT may provide a
more fundamental understanding of the sensitivity of the triple-alpha 
process to shifts in $\alpha_{\rm em}^{}$.


\section*{Acknowledgments}
We are grateful to Silas Beane and Martin Savage  for useful comments.
We thank Andreas Nogga for an updated
analysis of the $^4$He nucleus.  Partial financial support from Deutsche
Forschungsgemeinschaft and NSFC (Sino-German CRC 110), Helmholtz Association 
(contract VH-VI-417), BMBF\ (grant 06BN9006), and U.S. Department of
Energy (DE-FG02-03ER41260) is acknowledged. This work was further supported
by the EU HadronPhysics3 project, and by funds provided by the ERC project 
259218 NUCLEAREFT. Computational resources for this project were provided 
by the J\"{u}lich Supercomputing Centre (JSC) at the Forschungszentrum 
J\"{u}lich  and by RWTH Aachen.



\begin{thebibliography}{99}

\bibitem{1} 
  F.~Hoyle, 
  Astrophys.\ J.\ Suppl.\ Ser. \textbf{1}, 121 (1954).

\bibitem{2} 
  D.~N.~F.~Dunbar, R.~E.~Pixley, W.~A.~Wenzel, and W.~Whaling, 
  Phys.\ Rev. \textbf{92}, 649 (1953).

\bibitem{3} 
  C.~W.~Cook, W.~A.~Fowler, C.~C.~Lauritsen, and T.~Lauritsen,
  Phys.\ Rev. \textbf{107}, 508 (1957).

\bibitem{Linde} 
  A.~Linde, ``The inflationary multiverse,'' in {\it Universe or multiverse?}, edited by
  B.~Carr (Cambridge University Press, Cambridge, England, 2007).

\bibitem{Kragh} 
  H.~Kragh, 
  Arch.\ Hist.\ Exact Sci. \textbf{64}, 721 (2010).

\bibitem{6}
  B.~Carter, ``Large number coincidences and the anthropic principle'', in {\it Confrontation of cosmological theories with
  observational data}, edited by M.~S.~Longair (Reidel, Dordrecht, 1974).

\bibitem{7}
  B.~J.~Carr and M.~Rees, 
  Nature \textbf{278}, 605 (1979).

\bibitem{Weinberg:1987dv} 
  S.~Weinberg,
  Phys.\ Rev.\ Lett. {\bf 59}, 2607 (1987).

\bibitem{Susskind:2003kw} 
  L.~Susskind, ``The anthropic landscape of string theory,'' in {\it Universe or multiverse?}, 
  edited by B.~Carr (Cambridge University Press, Cambridge, England, 2007).

\bibitem{Livio}
  M.~Livio, D.~Hollowell, A.~Weiss, and J.~W.~Truran,
  Nature \textbf{340}, 281 (1989).

\bibitem{Oberhummer_astro} 
  H.~Schlattl, A.~Heger, H.~Oberhummer, T.~Rauscher, and A.~Cs\'ot\'o,
  Astrophys.\ Space Sci. {\bf 291}, 27 (2004).

\bibitem{Oberhummer:2000mn} 
  H.~Oberhummer, A.~Cs\'ot\'o, and H.~Schlattl,
  Nucl.\ Phys.\ A {\bf 689}, 269 (2001).

\bibitem{Oberhummer:2000zj}
  H.~Oberhummer, A.~Cs\'ot\'o, and H.~Schlattl,
  Science {\bf 289}, 88 (2000).

\bibitem{WeinbergFacing} 
  S.~Weinberg,
  ``Facing Up'' (Harvard University Press, Cambridge, Massachusetts, 2001).

\bibitem{Borasoy:2006qn} 
  B.~Borasoy, E.~Epelbaum, H.~Krebs, D.~Lee, and U.-G.~Mei{\ss}ner,
  Eur.\ Phys.\ J.\ A {\bf 31}, 105 (2007).

\bibitem{Dean_QMC}
  D.~Lee,
  Prog.\ Part.\ Nucl.\ Phys. {\bf 63}, 117 (2009).

\bibitem{Epelbaum:2009zsa} 
  E.~Epelbaum, H.~Krebs, D.~Lee, and U.-G.~Mei{\ss}ner,
  Eur.\ Phys.\ J.\ A {\bf 41}, 125 (2009).

\bibitem{Epelbaum:2009pd} 
  E.~Epelbaum, H.~Krebs, D.~Lee, and U.-G.~Mei{\ss}ner,
  Phys.\ Rev.\ Lett.\  {\bf 104}, 142501 (2010).

\bibitem{Epelbaum:2010xt} 
  E.~Epelbaum, H.~Krebs, D.~Lee, and U.-G.~Mei{\ss}ner,
  Eur.\ Phys.\ J.\ A {\bf 45}, 335 (2010).

\bibitem{Epelbaum:2011md} 
  E.~Epelbaum, H.~Krebs, D.~Lee, and U.-G.~Mei{\ss}ner,
  Phys.\ Rev.\ Lett.\  {\bf 106}, 192501 (2011).

\bibitem{Epelbaum:2008vj} 
  E.~Epelbaum, H.~Krebs, D.~Lee, and U.-G.~Mei{\ss}ner,
  Eur.\ Phys.\ J.\ A {\bf 40}, 199 (2009).

\bibitem{Epelbaum:2012qn} 
  E.~Epelbaum, H.~Krebs, T.~A.~L\"ahde, D.~Lee, and U.-G.~Mei{\ss}ner,
  Phys.\ Rev.\ Lett.\ {\bf 109}, 252501 (2012).

\bibitem{Epelbaum:2012iu} 
  E.~Epelbaum, H.~Krebs, T.~A.~L\"ahde, D.~Lee, and U.-G.~Mei{\ss}ner,
  Phys.\ Rev.\ Lett.\ {\bf 110}, 112502 (2013).

\bibitem{Oberhummer:1999ab} 
  H.~Oberhummer, A.~Cs\'ot\'o, and H.~Schlattl,
  arXiv:astro-ph/9908247.

\bibitem{Weinberg:1990rz} 
  S.~Weinberg,
  Phys.\ Lett.\ B {\bf 251}, 288 (1990).

\bibitem{Epelbaum:2008ga} 
  E.~Epelbaum, H.-W.~Hammer, and U.-G.~Mei{\ss}ner,
  Rev.\ Mod.\ Phys. {\bf 81}, 1773 (2009).

\bibitem{Machleidt:2011zz}
  R.~Machleidt and D.~R.~Entem,
  Phys.\ Rept.\  {\bf 503}, 1 (2011).

\bibitem{Berengut:2013nh} 
  J.~C.~Berengut, E.~Epelbaum, V.~V.~Flambaum, C.~Hanhart, U.-G.~Mei{\ss}ner, 
  J.~Nebreda, and J.~R.~Pel\'aez, 
  Phys.\ Rev.\ D {\bf 87}, 085018 (2013).

\bibitem{Mnucl}
  M.~Procura, B.~U.~Musch, T.~Wollenweber, T.~R.~Hemmert, and W.~Weise,
  Phys.\ Rev.\ D {\bf 73}, 114510 (2006).

\bibitem{Bernard:2007zu} 
  V.~Bernard,
  Prog.\ Part.\ Nucl.\ Phys. {\bf 60}, 82 (2008).

\bibitem{Colangelo}
  G.~Colangelo and S.~D\"urr, 
  Eur.\ Phys.\ J.\ C {\bf 33}, 543 (2004).

\bibitem{Gasser:1983yg} 
  J.~Gasser and H.~Leutwyler,
  Ann.\ Phys. {\bf 158}, 142 (1984).

\bibitem{Baron:2010bv}
  R.~Baron, P.~Boucaud, J.~Carbonell, A.~Deuzeman, V.~Drach, F.~Farchioni, 
  V.~Gimenez and G.~Herdoiza {\it et al.},
  JHEP {\bf 1006}, 111 (2010)

\bibitem{Bernard:2006te} 
  V.~Bernard and U.-G.~Mei{\ss}ner,
  Phys.\ Lett.\ B {\bf 639}, 278 (2006).


\bibitem{Luscher:1986pf} 
  M.~L\"uscher,
  Commun.\ Math.\ Phys. {\bf 105}, 153 (1986).

\bibitem{Luscher:1990ux} 
  M.~L\"uscher,
  Nucl.\ Phys.\ B {\bf 354}, 531 (1991).

\bibitem{Lee:2007a} 
  D.~Lee,
  Eur.\ Phys.\ J.\ A {\bf 35}, 171 (2008).

\bibitem{Nogga}
A.~Nogga, {\it private communication}. 

\bibitem{Bernard:1995dp} 
  V.~Bernard, N.~Kaiser, and U.-G.~Mei{\ss}ner,
  Int.\ J.\ Mod.\ Phys.\ E {\bf 4}, 193 (1995).

\bibitem{Buettiker:1999ap} 
  P.~B\"uttiker and U.-G.~Mei{\ss}ner,
  Nucl.\ Phys.\ A {\bf 668}, 97 (2000).


\bibitem{Epelbaum:2002gb} 
  E.~Epelbaum, U.-G.~Mei{\ss}ner, and W.~Gl\"ockle,
  Nucl.\ Phys.\ A {\bf 714}, 535 (2003).

\bibitem{Epelbaum:2002gk} 
  E.~Epelbaum, U.-G.~Mei{\ss}ner, and W.~Gl\"ockle,
  arXiv:nucl-th/0208040.

\bibitem{Beane:2002xf} 
  S.~R.~Beane and M.~J.~Savage,
  Nucl.\ Phys.\ A {\bf 717}, 91 (2003).

\bibitem{Beane:2006mx} 
  S.~R.~Beane, P.~F.~Bedaque, K.~Orginos, and M.~J.~Savage,
  Phys.\ Rev.\ Lett.\ {\bf 97}, 012001 (2006).

\bibitem{Beane:2001bc} 
  S.~R.~Beane, P.~F.~Bedaque, M.~J.~Savage, and U.~van Kolck,
  Nucl.\ Phys.\ A {\bf 700}, 377 (2002).

\bibitem{Chen:2010yt} 
  J.-W.~Chen, T.-K.~Lee, C.-P.~Liu, and Y.-S.~Liu,
  Phys.\ Rev.\ C {\bf 86}, 054001 (2012).

\bibitem{Soto:2011tb} 
  J.~Soto and J.~Tarrus,
  Phys.\ Rev.\ C {\bf 85}, 044001 (2012).

\bibitem{Epelbaum:2001fm} 
  E.~Epelbaum, U.-G.~Mei{\ss}ner, W.~Gl\"ockle, and C.~Elster,
  Phys.\ Rev.\ C {\bf 65}, 044001 (2002).

\bibitem{Epelbaum:2013ij} 
  E.~Epelbaum and J.~Gegelia,
  arXiv:1301.6134 [nucl-th].

\bibitem{Frink:2005ru} 
  M.~Frink, U.-G.~Mei{\ss}ner, and I.~Scheller,
  Eur.\ Phys.\ J.\ A {\bf 24}, 395 (2005).

\bibitem{Alarcon:2011zs} 
  J.~M.~Alarc\'on, J.~Martin Camalich, and J.~A.~Oller,
  Phys.\ Rev.\ D {\bf 85}, 051503 (2012).

\bibitem{Bedaque:2010hr} 
  P.~F.~Bedaque, T.~Luu, and L.~Platter,
  Phys.\ Rev.\ C {\bf 83}, 045803 (2011).

\end{thebibliography}
\end{document}